\documentclass[journal,10pt,doublecolumn]{IEEEtran}
\usepackage{multicol}
\usepackage{mathtools, cuted}
\usepackage{lipsum, color}

\usepackage{footnote}
\usepackage[table]{xcolor}
\usepackage{times}
\usepackage{enumitem}
\usepackage{epsfig}
\usepackage{amsmath}
\usepackage{amsthm}
\usepackage{amsfonts}
\usepackage{graphicx}
\usepackage{amssymb}
\usepackage{amstext}
\usepackage{latexsym}
\usepackage{color,colortbl}
\usepackage{ifthen}
\usepackage{verbatim}
\usepackage{array,tabularx}
\usepackage[mathscr]{euscript}
\usepackage{accents}
\usepackage{cite}
\usepackage{hhline}
\usepackage{caption}
\usepackage{subcaption}
\usepackage{xcolor}
\usepackage{mathtools}
\usepackage{url}
\usepackage{xparse}
\usepackage{makecell}
\usepackage{varwidth}
\usepackage{bm}
\usepackage[bookmarks,colorlinks]{hyperref}
\usepackage{arydshln}
\usepackage{footmisc} 

\usepackage{comment}
\usepackage{wrapfig}

\usepackage{float}
\usepackage{booktabs}

\usepackage{mathtools}

\usepackage{booktabs}

\usepackage{multirow}

\usepackage{acronym}

\newcolumntype{C}[1]{>{\centering\let\newline\\\arraybackslash\hspace{0pt}}m{#1}}

\usepackage[normalem]{ulem}

\usepackage[ruled, lined,linesnumbered, longend]{algorithm2e}
\usepackage{tikz}
\usetikzlibrary{fit, calc}

\colorlet{mygray}{gray!70}

\usepackage{algorithm}
\usepackage{algpseudocode}

\usepackage[ruled,linesnumbered]{algorithm2e}  
\SetKwInput{KwData}{\textbf{Init}} 
\SetKwInput{KwCurInput}{\textbf{Input}}

\theoremstyle{definition}

\theoremstyle{definition}

\theoremstyle{definition}

\newcommand{\off}[1]{}

\usetikzlibrary{matrix,decorations.pathreplacing}
\usetikzlibrary{arrows.meta}

\newcommand{\rev}[1] {\textcolor{black}{ #1}}

\newcommand{\mrev}[1] {\textcolor{black}{ #1}}

\definecolor{DarkGreen}{rgb}{0.1,0.5,0.1}
\definecolor{DarkRed}{rgb}{0.5,0.1,0.1}
\definecolor{DarkBlue}{rgb}{0.1,0.1,0.5}
\definecolor{DarkPurple}{rgb}{0.5,0.2,0.5}
\definecolor{DarkTurquoise}{rgb}{0.1,0.5,0.5}

\definecolor{beaublue}{rgb}{0.74, 0.83, 0.9}
\definecolor{coolblack}{rgb}{0.0, 0.18, 0.39}
\definecolor{apricot}{rgb}{0.98, 0.81, 0.69}
\definecolor{burntorange}{rgb}{0.8, 0.33, 0.0}
\definecolor{blue-violet}{rgb}{0.54, 0.17, 0.89}
\definecolor{byzantium}{rgb}{0.44, 0.16, 0.39}
\definecolor{brilliantrose}{rgb}{1.0, 0.33, 0.64}
\definecolor{cerisepink}{rgb}{0.93, 0.23, 0.51}
\definecolor{cobalt}{rgb}{0.0, 0.28, 0.67}
\definecolor{bostonuniversityred}{rgb}{0.8, 0.0, 0.0}

\newcommand{\loss}{\mathcal{L}}

\newcommand{\paren}[1]{\left( #1 \right)}

\usepackage{mathtools}

\acrodef{dnn}[DNN]{\rev{Deep Neural Network}}
\acrodef{ml}[ML]{\rev{Machine Learning}}
\acrodef{rtt}[RTT]{Round-Trip Time}
\acrodef{rnn}[RNN]{\rev{Recursive Neural Network}}
\acrodef{lstm}[LSTM]{\rev{Long Short-Term Memory}} 
\acrodef{ge}[GE]{Gilbert-Elliott}
\acrodef{snr}[SNR]{\rev{Signal to Noise Ratio}}
\acrodef{sinr}[SINR]{\rev{Signal to Interference and Noise Ratio}}
\acrodef{mse}[MSE]{\rev{Mean Squared Error}}
\acrodef{rmse}[RMSE]{\rev{Root Mean Squared Error}}
\acrodef{mcs}[MCS]{\mrev{Modulation and Coding Scheme}}
\acrodef{fc}[FC]{\rev{Fully Connected}}
\acrodef{fec}[FEC]{Forward Error Correction}
\acrodef{acrlnc}[AC-RLNC]{Adaptive and Causal Netwrok Coding}
\acrodef{phy}[PHY]{\rev{Physical}}

\DeclareMathOperator*{\argmax}{arg\,max}

\usepackage[all=normal,paragraphs=tight,floats=normal,mathspacing=normal,wordspacing=tight,charwidths=tight,mathdisplays=normal,leading=normal]{savetrees}

\ifCLASSINFOpdf
\else
\fi
\begin{document}
%
\title{\vspace{-0.4cm}DeepNP: Deep Learning-Based Noise Prediction\\ for Ultra-Reliable Low-Latency Communications\thanks{Parts of this work were presented at the IEEE International Symposium on Information Theory (ISIT) 2022 as the paper \cite{cohen2022deepnp}.}}
%
%
%

\author{%
   \IEEEauthorblockN{Adina Waxman\IEEEauthorrefmark{1},
                     Nir Shlezinger\IEEEauthorrefmark{2},
                     and Alejandro Cohen\IEEEauthorrefmark{1}}\\
   \IEEEauthorblockA{\IEEEauthorrefmark{1}%
                      Faculty of Electrical and Computer Engineering, Technion, Israel, adina.waxman@campus.technion.ac.il, alecohen@technion.ac.il}\\   
    \IEEEauthorblockA{\IEEEauthorrefmark{2}%
                     School of Electrical and Computer Engineering, Ben-Gurion University of the Negev, Israel, nirshl@bgu.ac.il}\\
\vspace{-0.8cm}}

\maketitle


\maketitle

\begin{abstract}
\rev{Adaptive network coding schemes provide a promising approach to bridging the gap between high data rates and low delay in real-time streaming applications. However, their effectiveness often relies on accurate channel prediction, which is typically based on delayed feedback and is especially challenging when the underlying channel model is unknown. To address this, we introduce a novel integration of network coding with a channel-agnostic, Deep learning-based Noise Prediction algorithm (DeepNP). Unlike traditional estimators, DeepNP predicts statistical noise rates rather than instantaneous noise realizations, significantly simplifying the prediction task while enhancing coding performance. DeepNP is designed to operate with both binary (e.g., acknowledgments) and continuous-valued (e.g., Signal-to-Noise Ratio, SNR) feedback. We incorporate DeepNP into the Adaptive and Causal Random Linear Network Coding (AC-RLNC) framework to jointly optimize throughput and in-order delivery delay. Two variants are proposed: (i) Erasure-Rate DeepNP (ER-DeepNP), which serves as a transport-layer noise predictor and achieves in a numerical study up to a 2× reduction in mean and maximum delay with less than 0.1 loss in throughput compared to statistic-based estimators, under Round-Trip Time (RTT) up to 40 time slots and erasure rates up to 60\%; and (ii) Cross-Layer DeepNP (CL-DeepNP), which dynamically adjusts the SNR threshold to maintain high physical layer code rates while achieving low transport-layer erasure rates. This yields, in the presented numerical study, a 25\% throughput gain over fixed-threshold approaches. Our results demonstrate that DeepNP enables robust, model-free noise prediction, making adaptive network coding more viable in practical, feedback-limited communication scenarios.}

\end{abstract}

  \section{Introduction}\label{sec:intro}
\mrev{Future wireless technologies support multiple service categories with distinct performance objectives. In particular, enhanced Mobile Broadband (eMBB) focuses on high data rates, whereas Ultra-Reliable Low-Latency Communications (URLLC) prioritizes extreme reliability and very low latency, typically requiring packet error probabilities on the order of $10^{-5}$–$10^{-7}$ and end-to-end latency below $1$~ms \cite{8329618,anand2018resource,dias2023sliding,vasudevan2026revisiting}.} However, simultaneously achieving \mrev{high reliability and low latency while maintaining efficient throughput} is often highly challenging. This stems from the fact that \mrev{reliable} high data rates are typically achieved by coding over large blocks, while low delay requires short blocks. As a result,  a trade-off is induced between throughput and in-order delay \mrev{under reliability guarantees}.
To mitigate the existing trade-off between \mrev{reliable} throughput and in-order delay, various coding schemes have been proposed in the literature  \cite{joshi2014effect,cloud2015coded,joshi2016efficient,yang2014deadline,TomFitLucPedSee2014,7117455,fong2019optimal,9076631,9245536,d2021post,emara2021low,9433517}.
\textcolor{black}{In this work, we focus on reliable real-time streaming under strict latency constraints, where adaptive coding mechanisms induce a trade-off between throughput efficiency and in-order delivery delay. Such settings are consistent with URLLC scenarios, where reliability is ensured by the coding mechanism while efficient transmission requires balancing retransmissions and delay.} Specifically, in the presence of delayed feedback, the works \cite{joshi2014effect,cloud2015coded,joshi2016efficient}  proposed codes to reduce the streaming delay over an erasure channel.
For coded blocks, the authors of \cite{yang2014deadline} proposed an adaptive solution, where the sender can choose the size of the next block and the number of information packets in the block for deadline-aware applications. 
The recently proposed \ac{acrlnc} scheme, applied to single-path, multi-path, and multi-hop networks  \cite{9076631,dias2023sliding,9245536,d2021post,esfahanizadeh2024benefits,cohen2022broadcast,vasudevan2026revisiting}, implements joint scheduling-coding in a manner that is both causal and adaptive. \mrev{Notably, \ac{acrlnc} enables reliable communication with zero decoding error probability \cite{9076631} and has been shown to achieve URLLC reliability targets with millisecond-scale latency \cite{dias2023sliding}.}
A core limitation of these adaptive coding methods as well as of existing classical coding solutions \cite{luby2002lt,shokrollahi2006raptor,LubMitShoSpiSte1997,KarLei2014,lieb2018complete} stems from the inherent challenge in providing reliable noise prediction. Specifically,  losses of throughput rate and high in-order delays often occur due to differences between the amount of noise the code was designed for and the actual realizations of the noise. 
As a result, to date, existing solutions are yet to close this trade-off and obtain the desired performance. For instance, while \ac{acrlnc} was shown to achieve over $90\%$ of the communication capacity, it often yields high in-order delay which is far from the optimal lower bound of the communication, due to inaccurate predictions of the noise variations.
Existing approaches estimate the average noise rate in some settings from delayed binary feedback via empirical estimates of the first and second-order moments~\cite{9076631,michel2022flec,dias2023millimeter}.
Moreover, while existing approaches rely on {\em binary} feedback, e.g., acknowledgments, in practice one may have access to channel quality indicators that are more informative than binary values, e.g., received \ac{snr}.
Particularly, the erasure pattern in the transport layer may be determined according to an \ac{snr} threshold determined by the \ac{mcs} \cite{dias2023sliding,dias2023millimeter, 10765762, CL}. These schemes are configured at the \ac{phy} layer to meet different performance requirements, as implemented in standards such as \cite{802_11ad_standard,schultz2013802,ccs_specs}. 
These overall observations motivate incorporating \ac{phy} layer noise prediction algorithms with transport layer adaptive network coding to implement complex prediction mappings exploiting continuous-valued feedback without relying on the channel model or its statistics.

Classical noise prediction methods like maximum-likelihood \cite{Du2011MaximumLB} and minimum-mean square-error \cite{8019834} require knowledge of the channel model or are computationally intensive as they require operations such as matrix multiplication and inversion. 
%
To predict noise in unknown conditions, deep learning approaches have gained significant attention. For example, \cite{8395053, 9143570, 8865038, 8052521} use learning methods to predict \ac{snr} and channel state information. However, while predicting the channel state is important for enhanced communication, focusing the learning process directly on the coding procedure can help the task at hand.
%
Works in this domain, such as \cite{emara2022ivory, song2025adaptive}, typically use reinforcement learning approaches that may not efficiently incorporate domain-specific knowledge about the system behavior and dynamics, potentially leading to longer convergence times and less interpretable solutions.
%
On top of that, the integration of transport layer transmission scheduling and \ac{phy} layer indications advocates for a cross-layer solution that aims to maximize total layers' rates.

Hence, our main contributions are as follows.  We develop an algorithm for predicting the noise realizations to narrow this significant gap in reliable streaming communication without relying on knowledge of the underlying channel model.  
To that aim, we propose a data-driven adaptive causal network coding for URLLC coined {\em DeepNP}.  
DeepNP augments network coding schemes with deep learning-based noise prediction, which is designed to learn from data the pattern of the communication link and predict the noise rate during the delayed feedback period - the \ac{rtt}.  
We present two variants of DeepNP: Erasure-Rate DeepNP (ER-DeepNP), which is utilized solely as a noise predictor, and Cross-Layer DeepNP (CL-DeepNP), which additionally adjusts the SNR threshold during transmission to achieve a higher throughput across both the physical and the transport layers. 

We design a dedicated \rev{\ac{dnn}}, whose architecture comprises an interconnection of neural building blocks with interpretable internal features representing the predictions for each block in an \rev{\acf{rtt}}. The \ac{dnn} is designed to process either binary (e.g., \rev{acknowledgments}) or continuous-valued (e.g., \ac{snr} values) delayed feedbacks. 
While accurately predicting the instantaneous noise realization in each time slot is exceptionally difficult, our design builds upon the insight that \emph{adaptive coding does not require these realizations}, and in fact relies on the noise rate, i.e., the rate of the errors induced by the channel during the delayed feedback period. Thus, we train the DNN in a manner that is invariant of the coding scheme, i.e., as a generative machine learning model~\cite{shlezinger2022discriminative}. Our training method boosts DeepNP to predict the noise rate while adding penalizing terms to encourage its internal features to represent predictions of the instantaneous noise. 
The resulting ER-DeepNP is integrated into the AC-RLNC coding scheme. The resulting method combines channel-model-based network coding with data-driven DNNs as a form of model-based deep learning \cite{shlezinger2020model,shlezinger2022model,shlezinger2023model}. Such designs have been shown to empower and robustify various communications algorithms \cite{ shlezinger2019viterbinet,shlezinger2019deepSIC,raviv2023adaptive, shlezinger2020inference}. 

For CL-DeepNP, we expand the ER-DeepNP design to predict the erasure rate over future time intervals longer than \ac{rtt}, adapting to varying \ac{snr} thresholds. This enhancement facilitates achieving a balance between the transport layer channel rate and \ac{phy} code rate.
We contrast the performance of the proposed approach with that of the channel-model-based \ac{acrlnc} \cite{9245536}, where the re-transmissions are scheduled at the sender using average statistical information. We show that the proposed DeepNP can gain up to a factor of $2$ in mean and maximum delay while maintaining a throughput difference of less than $0.1$. In CL-DeepNP, we demonstrate a $25\%$ increase in throughput with an adaptive threshold compared to a constant threshold approach.
This demonstrates that despite the inherent challenges of noise prediction, a properly designed and trained DNN-based noise predictor can still notably contribute to adaptive network coding. 

The structure of this work is as follows. In Section~\ref{sec:sys}, we formally describe the system model and the metrics in use and provide a background on adaptive causal network coding. In Section~\ref{sec:Deepnp} we present DeepNP and how it augments AC-RLNC. In Section~\ref{sec:evaluation}, we evaluate the performance of the proposed solution. Finally, we conclude the paper in Section~\ref{sec:conclusions}.

\section{System Model and Preliminaries}\label{sec:sys}
In this section, we present the system model and review the preliminaries needed for our derivation of DeepNP. We first introduce the considered communication model in Subsection~\ref{sec:sys_channel} and the problem formulation in Subsection~\ref{subsec:problem}. Following that, we review the relevant background in adaptive and causal network coding in Subsection~\ref{subsec:Preliminaries}. 

\subsection{Communication Channel Model}\label{sec:sys_channel}
We consider a point-to-point real-time slotted communication system with delayed feedback. At each time slot $t$, the sender transmits a coded packet $c_t$ to the receiver over a single-path noisy forward channel. 

The noisy forward channel may erase packets according to, e.g., low instantaneous \ac{snr}. The instantaneous erasure probability at time slot $t$ is denoted as $p_t$, while the erasure rate is denoted $\epsilon$, exhibiting the relationship:
\begin{equation*}
    \epsilon = \mathop{\lim}_{T \rightarrow \infty} \frac{1}{T}\sum_{t=1}^{T}p_t.
\end{equation*}

The corresponding channel rate $r$ is given as:
\begin{equation}
    r = 1 - \epsilon.
\end{equation}

Here, unlike classical models and solutions considered in the literature \cite{cover2012elements}, we assume the channel model and its statistics are unknown to the sender and the receiver. However, the sender can track the channel statistics by delayed feedback and predict the next channel realizations.

We assume the arrival of a packet to the receiver is acknowledged by the sender over a noiseless feedback channel\footnote{To focus on the adaptive coding aspect of transmission, we assume that the feedback channel is noiseless. However, for a noisy feedback channel, we can consider techniques of cumulative feedback, e.g., as given in \cite{malak2019tiny}.}.
The time delay between the transmission of a packet and the time its corresponding feedback message arrives back at the transmitter is denoted {\em \acf{rtt}}. The transmission delay of a coded packet in bits/seconds is denoted by $t_d$, and the maximum propagation delay is denoted by $t_{\text{prop}}$. We assume that the size of the feedback acknowledgment is negligible, and fix the propagation delay for transmitted coded packets. The \ac{rtt} for each coded packet is ${\rm RTT} = t_d + 2t_{\text{prop}}$. For simplicity, we assume that the propagation delay in both the forward and feedback channels is $\frac{\rm RTT}{2}$. Therefore, if a coded packet $c_t$ is sent at time $t$, its corresponding feedback message, denoted by $F_t$, arrives at the sender at time $t+\rm RTT$.

Thus, using an adaptive and casual network code $\mathcal{C}$, at each time $t$,  some raw data packets $\{P_i\}_{i\leq t}$ are mapped and transmitted as a codeword $c_t$ based on the received feedback, such that
\begin{equation}
    c_t = \mathcal{C}\Big(\{P_i\}_{i\leq t},  \{F_i\}_{i\leq t^-}\Big),
    \label{eqn:NetCode}
\end{equation}
where we define
\begin{equation}\label{eq:t-}
    \quad t^-\triangleq t-{\rm RTT}.
\end{equation}
While the transmission is considered to take place in the transport layer, we suggest two feedback mechanisms: one that remains in the domain of the transport layer and one that takes advantage of the \ac{phy} layer indications. With $b_t$ being the binary acknowledgments corresponding to the feedback message $F_t$, we consider the following:

\begin{itemize}[wide, labelwidth=0.3cm, labelindent=1pt]
    \item {\em Binary feedback}: \rev{The sender receives binary Acknowledgement (ACK) and Negative-Acknowledgement (NACK), with  $b_t=F_{t}=1$ signifying successful packet reception and $b_t=F_{t}=0$ denoting packet erasure, respectively.} 

    \item {\em Continuous feedback}: For each packet $c_t$, the sender receives the \ac{snr} level measured at the receiver during time slot $t+\frac{\rm RTT}{2}$, denoted as $s_t$. The feedback is, therefore, $F_t = s_t$, and the conversion from \ac{snr} level to binary ACK and NACK are obtained by a threshold parameter $\rm S_t^{\rm th}$, similar to \cite{10765762, CL}, where the erasure rate is defined as a function of the \ac{snr}.
    Thus, at time $t$,  the sender receives feedback $F_{t^-}$  indicating the reception status of coded packet $c_{t^-}$, and the conversion from \ac{snr} level to binary erasures is:
\begin{equation}
    \label{eqn:b_t}
    b_{t^-} \triangleq \begin{cases}
        1 \text{ (i.e., ACK for } c_{t^-}), & F_{t^-} \geq \rm S_{t^-}^{\rm th}\\
        0 \text{ (i.e., NACK for time } c_{t^-}), & \text{otherwise}.
    \end{cases}
\end{equation}
 In practice, the \ac{snr} feedback is conveyed in a high-resolution digital representation, which can be treated as a continuous range for design purposes.
\end{itemize}
The threshold $\rm S_t^{\rm th}$ can be selected according to the \ac{mcs} used in the \ac{phy} layer of the communication standards to evaluate the real-world systems \cite{802_11ad_standard,schultz2013802,ccs_specs}.
Thus, $\rm S_t^{\rm th}$ determines both the transport layer erasure pattern and rate, as well as the \ac{phy} layer code rate. To formalize this, let $R_t^{\rm Phy}$ be the code rate at which coded packet $c_t$ is transmitted at the \ac{phy}  layer. The code rate $R_t^{\rm Phy}$ is uniquely determined by $\rm S_t^{\rm th}$ through the  \ac{mcs}. Typically, the code rate setting is handled using a look-up table based on the \ac{snr} limit, e.g., as illustrated in Table~\ref{tab:mcs}. 
Note that tuning $\rm S_t^{\rm th}$ to determine both the transport erasure pattern and the \ac{phy} code rate is also applicable for binary feedback.
\begin{table}[]
    \centering
    \begin{tabular}{|c|c|} \hline  
      Code Rate&SINR limit [dB]\\ \hline  
          
     1/2&1\\ \hline  
  5/8&5\\ \hline  
  3/4&8\\\hline
  13/16&12\\\hline \end{tabular}
\caption{MCS code rate per SNR limits.}
\label{tab:mcs}
\vspace{-0.6cm}
\end{table}

\subsection{Problem Formulation}
\label{subsec:problem}

Our goal is to derive an adaptive coding scheme that forms encoded raw data packets into encoded packets, based on past feedback information, aiming to maximize throughput and minimize the in-order delivery delay. 

\subsubsection{Metrics}
To evaluate the performance of our solution we use the metrics defined as follows.

\noindent (1) {\bf Throughput, $\eta$}. This is the total amount of information data, in units of bits per second, arrived at the receiver in the transport layer. This measure is typically evaluated based on the following normalized forms:

\smallskip
\noindent (1.1) {\bf Normalized Throughput, $\eta_{N}$} This is the total amount of information data arrived at the receiver divided by the total amount of bits transmitted by the sender, in the transport layer.
Since we consider a slotted transmission with a fixed number of $T$ slots, with one packet transmitted at each slot and a constant packet size, the normalized throughput can be calculated as:
\begin{equation}
    \eta_{N}(\mathcal{C}) \triangleq \frac{d(\mathcal{C})}{T- \frac{\rm RTT}{2}},
    \label{eqn:eta_n}
\end{equation}
where $d(\mathcal{C})$ is the number of packets successfully decoded at the receiver using network code $\mathcal{C}$.

\noindent (1.2) {\bf Joint Normalized Throughput, $\eta_{J}$} This is the normalized throughput factored by the \ac{phy} layer code rate. The mean code rate at the \ac{phy} layer of all transmitted packets is, therefore,
\begin{equation*}
    \overline{R}^{\rm Phy}(\mathcal{C}) = \frac{1}{T} \sum_{t=0}^{T} R_t^{\rm Phy}(\mathcal{C}).
\end{equation*}
Hence, the Joint Normalized Throughput is given by, 
\begin{equation}
\label{eqn:eta_j}
    \eta_{J}(\mathcal{C}) = \eta_{N}(\mathcal{C}) \cdot \overline{R}^{\rm Phy}(\mathcal{C}).
\end{equation}
This metric captures the tradeoff between the \ac{phy} layer code rate and the transport layer erasure rate reflected in the transport layer throughput.

\smallskip
\noindent (2) {\bf In-Order Delivery Delay, $D_i$}. This is the difference between the time slot in which an information packet is first transmitted by the sender and the time slot in which it is decoded in order by the receiver. 
Specifically,  $T_1(P_i; \mathcal{C})$ is the first time packet $P_i$ is transmitted in any encoded packet of network code $\mathcal{C}$ and $T_d(P_i; \mathcal{C})$ is the time packet $p_i$ is decoded at the receiver. Therefore, the in-order delivery delay of each packet of index $i$ is:

\begin{equation}
    D_i(\mathcal{C}) \triangleq T_d(P_i; \mathcal{C}) - T_1(P_i; \mathcal{C}).
    \label{eqn:delay_p}
\end{equation}
The delay measure given in \eqref{eqn:delay_p} varies per packet. Thus, to determine the overall performance of the network, we give two different measures for delay evaluation: 

(2.1) {\bf Mean Delay, $D^{\text{mean}}$} - The {\em average} delay of all information packets sent in the network, i.e.,
\begin{subequations}
\label{eqn:D_mean_max}
    \begin{equation}
    D^{\text{mean}}(\mathcal{C}) \triangleq  \frac{1}{d(\mathcal{C})}\sum_{i=1}^{d(\mathcal{C})}D_i(\mathcal{C}).
    \label{eqn:D_mean}
    \end{equation}
This measure represents the network’s ability to send the entirety of the data.

(2.2) {\bf Maximum Delay, $D^{\text{max}}(\mathcal{C})$} - The {\em maximum} delay of all packets sent in the network, namely,
\begin{equation}
    D^{\text{max}}(\mathcal{C}) \triangleq \max_{i\in\{1,\ldots, d(\mathcal{C})\}} D_i(\mathcal{C})
   . \label{eqn:D_max}
    \end{equation}
\end{subequations}
This quantity represents the ability of the network to send information quickly in a real-time streaming scenario.

\subsubsection{Objective}
Based on the above metrics, we aim to design a coding scheme that maximizes the throughput within some time horizon $T$, while complying with predefined limits on the delay figures-of-merit. We differentiate between two goals: one is focused solely on achieving the throughput-delay tradeoff in the transport layer, and one that aims at high throughput across both layers.

Accordingly, the first design problem is formulated as
\begin{align}
\label{eqn:ProbForm_T}
    \mathop{\argmax}\limits_{\mathcal{C}} \,\,&\eta_{N}(\mathcal{C}) \\
    \text {subject to  } &  D^{\text{mean}}(\mathcal{C}) \leq \bar{D}_{\text{mean}} \notag \\
    &  D^{\text{max}}(\mathcal{C}) \leq \bar{D}_{\text{max}}, \notag 
\end{align}
where $\bar{D}_{\text{mean}}$ and $\bar{D}_{\text{max}}$ are the limitations imposed on the mean and max delay, respectively. For design purposes, the sender has access to data comprised of past transmissions and their corresponding feedback taken from the channel.
The second problem is similar, with the joint throughput as the objective
\begin{align}
\label{eqn:ProbForm_CL}
    \mathop{\argmax}\limits_{\mathcal{C}} \,\,&\eta_{J}(\mathcal{C}) \\
    \text {subject to  } &  D^{\text{mean}}(\mathcal{C}) \leq \bar{D}_{\text{mean}} \notag \\
    &  D^{\text{max}}(\mathcal{C}) \leq \bar{D}_{\text{max}}, \notag 
\end{align}

\subsubsection{Challenges}
An adaptive network code as in \eqref{eqn:NetCode} can be designed in light of the figures-of-merit in \eqref{eqn:ProbForm_T} and \eqref{eqn:ProbForm_CL} by encoding redundant information in the form of re-transmissions of the raw data packets, $\{p_i\}_{i\leq T}$. Based on this operation, we identify the following core challenges in designing adaptive network codes as in \eqref{eqn:NetCode}:

\begin{enumerate}[wide, labelwidth=0.3cm, labelindent=1pt, label={C\arabic*}]
    
    \item \label{itm:tradeoff_T} 
    {\em Throughput - delay tradeoff} - Re-transmissions of packets induce an inherent tradeoff between the throughput and in-order delay. While excessive re-transmissions reduce throughput (by decreasing $d(\mathcal{C})$), they also increase delay (by increasing $D_i(\mathcal{C})$), and vice versa.
    \item \label{itm:Domain} 
    {\em Channel modeling} - The design of the adaptive network coding schemes requires statistical knowledge of the channel behavior to determine the instantaneous coding procedure. Thus, the \ac{rtt} delay period in the feedback channel necessitates some form of prediction mechanism. In practice, such knowledge is typically approximated and possibly unfaithful to the actual underlying channel.
    Since statistical-based estimators assume a known channel model, their predictions tend to be inaccurate when such knowledge is unavailable or mismatched. Thus, predicting the instantaneous noise realization is often challenging.
    \item \label{itm:snr} 
    {\em Exploiting non-binary feedback} - existing non-coded \cite{weldon1982improved,anagnostou1986performance,ausavapattanakun2007analysis} and network coding schemes in the higher layers (e.g., transport layer) are typically designed to process binary feedback, i.e., ACK/NACK. As in practice, one often has access to \ac{snr} estimates. Such information is not readily incorporated into existing encoders in higher layers.  
    \item \label{itm:tradeoff_CL}  {\em Cross-LayerThroughput} - The erasure pattern in the transport layer may be determined by the \ac{mcs} in the \ac{phy} layer. This obtains a tradeoff between the \ac{phy} layer code rate and the transport layer erasure rate. While the presence of non-binary feedback allows setting the \ac{phy} layer modulation in a cross-layer manner, such joint designs are non-trivial. A modulation with a higher \ac{snr} threshold will result in a higher \ac{phy} layer code rate, but also with a higher erasure rate at the transport layer, and vice-versa. Balancing these two is beneficial to achieve a higher throughput over the entire network.
    
\end{enumerate} 

While Challenge \ref{itm:tradeoff_T} is shared by most sliding window network coding techniques \cite{cloud2015coded,9076631}, the consideration of \ref{itm:Domain} and \ref{itm:snr} is much less treated in the literature. Here, we aim to jointly tackle \ref{itm:tradeoff_T}-\ref{itm:snr} while deviating from conventional layer-wise separation into a cross-layer design~\ref{itm:tradeoff_T} by leveraging the availability of data via a {\em learning-aided design}, as detailed in Section~\ref{sec:Deepnp}.

\subsection{Preliminaries}
\label{subsec:Preliminaries}

\begin{figure}
\centering
  \includegraphics[width = 0.4 \textwidth, keepaspectratio]
  {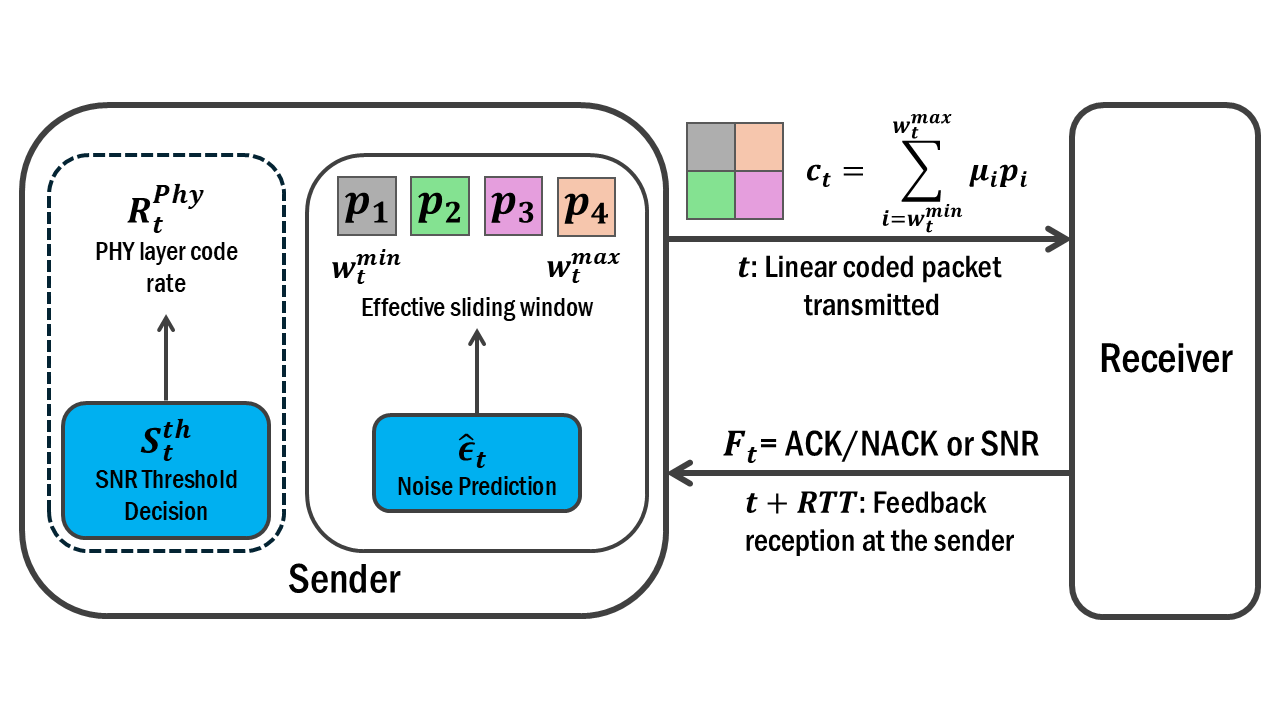}
  \vspace{-0.4cm}
  \caption{System Model. The sender uses the feedback to estimate the channel rate and encodes packets using \ac{acrlnc}. It may adjust the \ac{phy} code rate for the next transmission.}
  \label{fig:system_model}
  \vspace{-0.4cm}
\end{figure}

Our design detailed in the next section builds upon the \ac{acrlnc} scheme proposed in \cite{9076631}, which implements adaptive and causal network coding. Thus, we next briefly recall the \ac{acrlnc} scheme, whose model is illustrated Fig.~\ref{fig:system_model}. There, the encoding process carried out by the sender realizes the considered form of rateless adaptive and causal network coding.

\ac{acrlnc} uses a sliding window version of RLNC to code information packets into linear combinations \cite{lun2008coding, slw2013}. Let $\mu_i$ and $P_i$ denote the random coefficients drawn from a sufficiently large field and the raw information packets, respectively. The coded linear combination transmitted called a \rev{Degree of Freedom (DoF)}, is given by
\begin{equation}
\label{eqn:dof}
    c_{t}=\sum_{i=w_t^{\min}}^{w_t^{\max}}\mu_{i}P_{i},
\end{equation}
where $w_t^{\min}$  and  $w_t^{\max}$  are the beginning and end of the sliding window at time $t$, respectively.

To cope with channel erasures, the sender tracks the feedback and decides at each time step whether to transmit a new coded linear combination or re-transmit the same coded packet. Here, “new” and “same” refer to the raw information packets contained in the linear combination. Thus, re-transmitting the same coded packet means the same raw information packets are sent but with different random coefficients. In other words, $w_t^{\max} = w_{t-1}^{\max}$. For a new coded packet, a new raw information packet is added to the combination and so $w_t^{\max} = w_{t-1}^{\max}+1$. Note that $w_t^{\min}$ is incremented to the oldest raw packet sent that has not yet been decoded, as indicated by the feedback acknowledgments.

The sender uses two \acf{fec} mechanisms, a prior and a posterior, to counteract the erasure. 
The a priori mechanism transmits $\lceil\epsilon\cdot k\rfloor$ repeated DoFs, with $\lceil \cdot \rfloor$ denoting rounding to the nearest integer, after every $k$ transmissions of new information packets.
 In the posterior mechanism, at time $t$, the sender compares the channel rate $r$ with the DoF rate, that is, the ratio between missing new packets, $md_t$, to received re-transmissions, $ad_t$. When the channel rate is lower than the rate of the DoFs,  $d \triangleq \frac{md_t}{ad_t}$, the decoder does not have sufficient DoFs to decode the delivered packets, and a re-transmission is suggested. 

The \ac{fec} mechanisms require the sender to know the channel rate $r$ and the corresponding erasure rate $\epsilon$.
However, the sender cannot compute these quantities exactly due to the \ac{rtt} delay. At time step $t$, it can only compute these quantities for time step $t^{-}$ (defined in \eqref{eq:t-}), using the delayed feedback. Hence, with a tuneable parameter $th_p$, the DoF rate gap is given by  
\begin{equation*}
    \Delta_t \triangleq \frac{md_{t^-}}{ad_{t^-}}-r - th_p,
\end{equation*}
and retransmission is suggested at each time step for which 
\begin{equation*}
    \Delta_t>0,
\end{equation*} 
where $th_p$ is used to set the desired tradeoff between the throughput and delay (with a higher threshold value corresponding to higher throughput and higher delay, and the opposite).

In practice, $r$ and $\epsilon$ are not known to the sender. However, the server has access to the erasure noise realizations for times before $t^-$. This can be utilized to estimate the erasure rate at the unknown period of $[t^-, t]$ denoted by $\epsilon_t$, defined as
\begin{equation}
\label{eqn:SumSuccess}
    \epsilon_{t} \triangleq 1 - \frac{1}{\rm RTT} \sum_{j=t^-+1}^{t}b_j.
\end{equation}
The erasure rate dictates channel rate via $r_t = 1 - \epsilon_t$.
For instance, such a  statistic-based estimation of the erasure rate, denoted $\hat{\epsilon}_{t}$,  can be calculated via 
\begin{equation}\label{e_est}
    \hat{\epsilon}_{t} = 1 - \frac{\sum_{j=1}^{t^{-}} b_j}{t^{-}} + \frac{\sqrt{V}}{\rm RTT},
\end{equation}
where $V$ is the variance of the channel during the period of \ac{rtt}. We refer the readers to \cite{9076631,9245536} for detailed examples of how the tracked quantities and estimation presented above are computed based on channel modeling. 

Finally, to manage the maximum delay, a maximum sliding window of size $w$ is defined, such that $w_t^{\max}-w_t^{\min}\leq w$. When this limit is reached, the sender transmits the same packet until all the information packets in the linear combination transmitted are decoded. We again refer the readers to \cite{9076631,9245536} for further details on the operation of \ac{acrlnc}.

While \ac{acrlnc} can achieve improved throughput-delay trade-off by adapting the required re-transmissions, its adaptation relies on tracking the channel, e.g., the erasure probability $\epsilon_t$. However, when the channel exhibits high variations in its conditions,  statistic-based estimators as in \eqref{e_est} are likely to be inaccurate as noted in \ref{itm:Domain}. This, in turn, results in too few or too many re-transmissions, affecting the throughput-delay tradeoff in \ref{itm:tradeoff_T}.  
Considering \ac{snr} values to represent the channel conditions in the transport layer presented in \ref{itm:snr} is a quite new approach, leading to the question of the cross-layer throughput presented in \ref{itm:tradeoff_CL}. To better estimate the channel behavior and close these gaps, we propose a data-driven approach, augmenting network-coding-based protocols with a dedicated \ac{dnn} as described next.

\section{Deep Learning Aided Noise Prediction}
\label{sec:Deepnp}
In this section, we derive a scheme to better estimate the noise rate necessary for adaptive network coding.
Additionally, we propose an \ac{snr} adjusting scheme to optimally balance the PHY code rate and the transport erasure rate.
The solution proposed in this section comprises a \ac{dnn}-aided noise prediction module, coined {\em DeepNP}, integrated into an adaptive network code.
The complex and unknown underlying statistical relationship (\ref{itm:Domain}) motivates the data-driven approach that relies on deep learning methods, known for their ability to disentangle semantic information in complex settings. 
While predicting the instantaneous noise realization is highly challenging, our design is based on the key insight that adaptive coding solutions do not require these exact realizations, but rather use the overall noise \rev{rate. 
Hence,} we use a \ac{dnn} to predict noise rate based on the available feedback at each slot $t$. We use DeepNP to realize two forms of augmentations of an adaptive network code: as an {\em erasure rate} predictor, and for a {\em cross-layer} design that adjusts the erasure rate along with the \ac{snr} threshold. 

DeepNP predicts the noise rate during the unknown \ac{rtt} period. It aims to improve the statistical estimation of the erasure rate $\epsilon_t$ defined in \eqref{eqn:SumSuccess}.
The \ac{phy} layer indications may facilitate the learning process, by serving a continuous representation of the channel conditions. The memory of the channel, if it exists, further enhances the performance as well. The architecture and training procedure of DeepNP are presented in Subsections \ref{subsec:DeepNPBasic_Arch} and \ref{subsec:DeepNPBasic_Train}
respectively. Our first network code, termed \rev{{\em Erasure Rate DeepNP (ER-DeepNP)}}, augments DeepNP as an error predictor with the \ac{acrlnc} scheme, as detailed in Subsection~\ref{subsec:DeepNPBasic_acrlnc}.
ER-DeepNP is readily extended into \rev{{\em Cross-Layer DeepNP (CL-DeepNP)}}, designed to meet the cross-layer throughput goal \ref{itm:tradeoff_T}. While the packets during times $[t^-, t-1]$ are already transmitted at a certain code rate in the \ac{phy} layer, the code rate for the packets coming afterward is alterable. Thus, the CL-DeepNP predicts the noise rate for longer periods with a varying \ac{snr} threshold to optimally tune the \ac{phy} code rate with the transport layer erasure rate.
CL-DeepNP is described in Subsection~\ref{subsec:DeepNPExtended}, followed by a discussion in Subsection~\ref{subsec:disc}.

\begin{figure}
    \centering
    \begin{subfigure}{0.75\linewidth}
        \includegraphics[width=\linewidth,keepaspectratio]
        {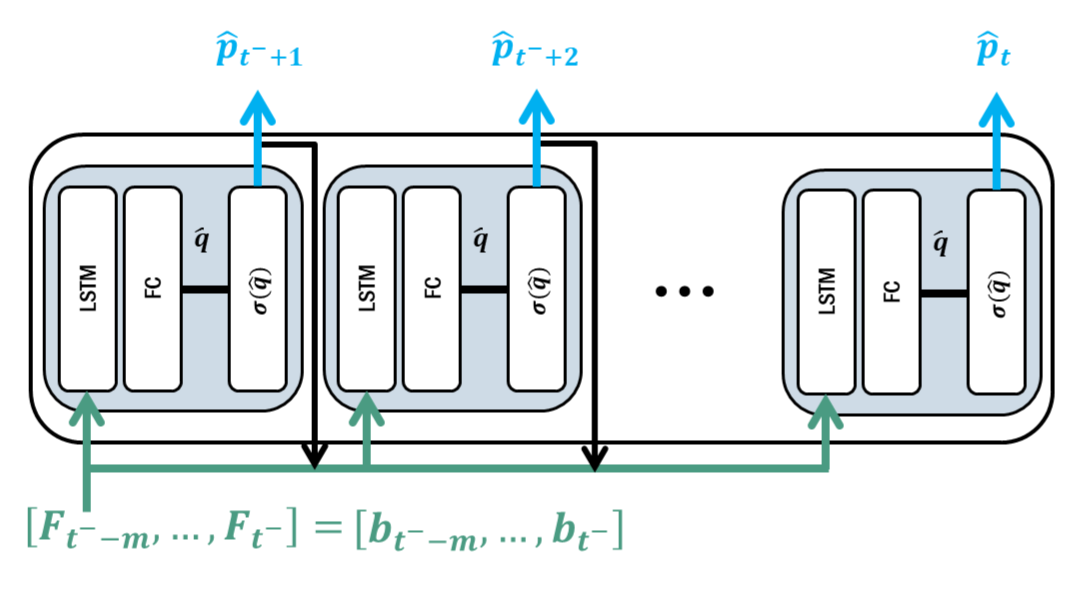}
        \caption{Binary feedback predictor}
        \label{fig:arch_dnn_bin}
    \end{subfigure}
    \begin{subfigure}{0.75\linewidth}
        \includegraphics[width=\linewidth,keepaspectratio]
        {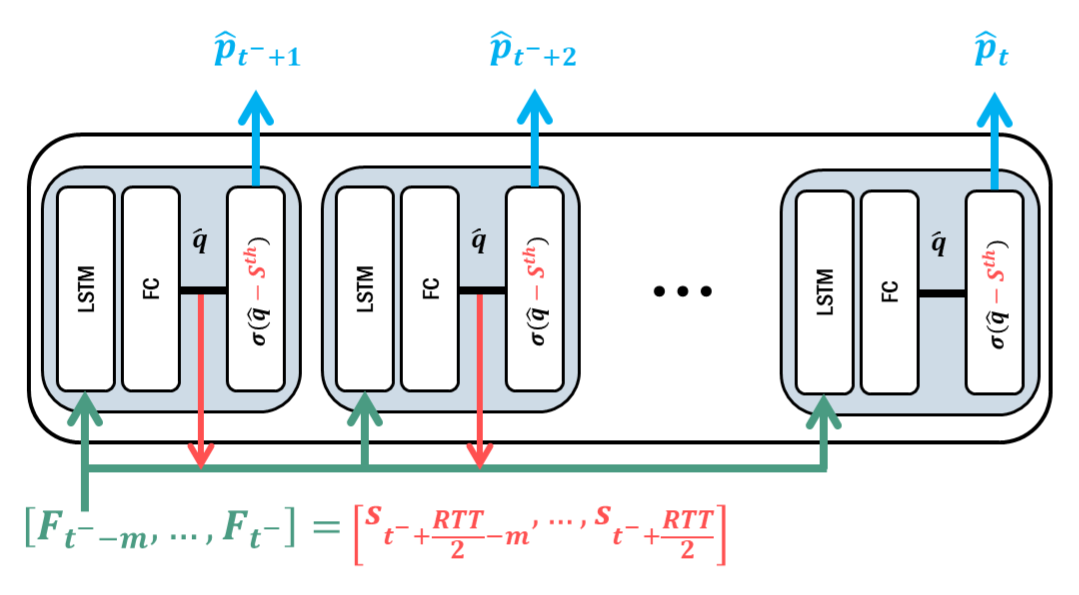}
        \caption{\ac{snr} feedback predictor}
        \label{fig:arch_dnn_snr}
    \end{subfigure}
    \caption{DeepNP architectures for binary and \ac{snr} feedback types. (a) For the binary feedback type, each block's input includes the output erasure probability. (b) For the \ac{snr} feedback type, the feedback is concatenated with the predicted \ac{snr} (i.e., the output of the fully-connected layer).}   
    \label{fig:DeepNP_architecture}
\end{figure}

\subsection{DeepNP - Architecture}
\label{subsec:DeepNPBasic_Arch}
The DeepNP architecture is designed to aid adaptive network coding by mapping a set of past feedback values into a soft estimate of the instantaneous erasures within an \ac{rtt} period. 
We propose a trainable architecture based on \ac{lstm} cells \cite{hochreiter1997long}. 
In order to leverage architectures that are capable of faithfully processing time sequences in a relatively low complexity manner, we opt the usage of RNN architectures, and particularly LSTM cells. In general, the proposed architecture can be enhanced to alternative RNN based solutions such as GRUs, state-space models~\cite{gu2021efficiently}, self-attention-based models~\cite{vaswani2017attention}, and other deep learning models for time sequence processing~\cite{shlezinger2024ai}. Nonetheless, in our numerical study, we choose LSTM cells as the basic architecture our DeepNP model, owing to its simplicity and ability to train with  limited data.
While in general noise prediction is statistically related to all past feedback, here we fix the number of past feedback used as input for noise prediction, and the internal memory of the \ac{lstm} units is exploited to learn longer-term correlations. Letting $m$ denote the hyperparameter representing the number of past feedback considered. The architecture maps the feedback $[F_{t^--m+1} \ldots F_{t^-}]$ into a set of predicted instantaneous erasure probabilities for the subsequent unknown \ac{rtt} slots, denoted $\{\hat{p}_j\}_{j=t^-+1}^t$.

Our architecture is based on the assumption that adjacent channel realizations are likely to be more correlated. Hence, the prediction at time instance $t$ is expected to be facilitated if provided with the prediction at time $t-1$. Accordingly, we design DeepNP as an interconnection of \ac{rtt} neural building blocks~\cite{shlezinger2020model}.  The $j$th building block serves to estimate a single probability $\hat{p}_j$ based on the feedback values and the preceding neural building blocks outputs. 
Each building block consists of an \ac{lstm} layer, followed by a \ac{fc} layer and a sigmoid function, $\sigma(\cdot)$.  A schematic of the architecture is depicted in Fig.~\ref{fig:DeepNP_architecture}. 
For the binary feedback case, the input to the $j$th block aggregates the estimations $\{\hat{p}_l\}_{l=t^-+1}^{j-1}$ obtained from the preceding blocks. Thus, the $j$th block input is: $[F_{t^--m+1} \ldots F_{t^-},\hat{p}_{t^-}+1 \ldots \hat{p_t}]$.
For continuous feedback, two alterations are made: $(i)$  to translate \ac{snr} to erasure rate, the \ac{fc} layer output, denoted $\hat{q_j}$,  is shifted by the \ac{snr} threshold $\rm S_t^{\rm th}$ before entering the sigmoid function. Meaning,  $\hat{p}_j =\sigma(\hat{q_j} - {\rm S_t^{\rm th}})$;
$(ii)$  to match the input units, the $j$th block input aggregates the instantaneous \ac{snr} prediction instead of the predicted erasure probabilities, meaning, the $j$th block input is $[F_{t^--m+1} \ldots F_{t^-},\hat{q}_{t^-}+1 \ldots \hat{q_t}]$. 

\subsection{DeepNP - Training}
\label{subsec:DeepNPBasic_Train}
DeepNP is trained in a generative manner, namely, separately from the network coding task~\cite{shlezinger2022discriminative}. This allows the same trained \ac{dnn} to be readily incorporated into different network coding mechanisms (and in general, for coding-based schemes). To describe its training procedure, we first formulate the data used for learning, then describe the loss function on which the training is carried out. 

\subsubsection{Data} \label{sec:data_basic}
To train the architecture detailed in Subsection~\ref{subsec:DeepNPBasic_Arch}, we utilize the access to past channel feedback to form a labeled dataset. Particularly, the dataset, $\mathcal{D}$, comprises a set of observed $m$ feedback values and their corresponding erasure values as labels:
\begin{equation*}
    \mathcal{D} = \{\{F_l\}_{l=t^--m+1}^{t^-}, \{b_j\}_{j=t^-+1}^t\}.
\end{equation*}
Note that when in the continuous feedback case, the binary values $\{b_j\}$ can be obtained from their continuous counterparts via \eqref{eqn:b_t}, where $S_t^{\text {th}}$ is predetermined and constant for any $t$.

\subsubsection{Loss} \label{sec:Loss_basic
} 
The learning process is conducted in a supervised manner, relying on two main loss components, detailed next. They are combined into a single loss measure used for training via conventional stochastic gradient-based learning:

\noindent (1) {\em Mean-Loss:}
To get an accurate estimation of the mean erasure rate over the \ac{rtt} period we use the \ac{rmse} between predicted erasure rates and the actual erasures. To boost successful individual predictions, we use the cross entropy loss, ${\rm CE}(\cdot;\cdot)$, on each block output separately. We assign larger weights to earlier time slots, as our neural building block architecture exploits earlier predictions to facilitate future ones. This is achieved using logarithmic weight decay as in \cite{samuel2019learning,lavi2023learn}, encouraging the DNN to be more confident in learning them. The resulting loss is, therefore,
\begin{align}
    &\loss_M\paren{\{\hat{p}_j\}_{j=t^-+1}^t,\{b_j\}_{j=t^-+1}^t} =\paren{\sum_{j=t^-+1}^{t} \paren{\hat{p}_j - b_t}^{2}}^{1/2} 
    \notag \\ &\qquad + \lambda\sum_{j=t^-+1}^{t} \log \paren{t-j+1}{\rm CE}\paren{\hat{p}_j,b_j},
    \label{eqn:Loss_M}
\end{align}
where $\lambda > 0$ is a regularization hyperparameter.

\smallskip
\noindent (2) {\em Bhattacharyya-Loss:}
To further boost the prediction accuracy, we use the Bhattacharyya Distance given by \cite{shannon1967lower,viterbi2013principles,dalai2014elias,barg2005distance}, a similarity measure for probability distributions. 
The Bhattacharyya Distance in the general case is defined as:
\begin{equation*}
    l_B(c, c') = -\ln(BC(c, c')),
\end{equation*}
with $BC$ being the Bhattacharyya Coefficient defined as:
\begin{equation*}
    BC(c, c') = \sum_y \sqrt{W(y|c)W(y|c')},
\end{equation*}
with $W(y|c)$, $W(y|c')$ being the channel transition probabilities from inputs $c$ and $c'$ to output vector $y$, respectively.

As shown in \cite{9076631}, the Bhattacharyya Distance establishes an upper bound for the throughput $\eta$, via
\begin{equation*}
\eta \leq \hat{r_t} - l_B(\hat{r_t}, r_t).
\end{equation*}
 As the equation's right side approaches 0, the predicted rate gets closer to the true rate, implying a potential for the throughput to approach the channel rate.

 In practice, $l(\hat{r_t}, r_t)$ is calculated as if the channel was a binary erasure channel which serves as an upper bound to our case and is readily  implemented as,
 \begin{equation*}
   \begin{split}
      BC(r_t, \hat{r}_t) = & \sum_{t=0}^{\rm RTT} \binom{\rm RTT}{t} \\
      & \times \left( ((r_t)(\hat{r}_t))^{t/2} \cdot ((1-r_t)(1-\hat{r}_t))^{\frac{\rm RTT -t}{2}} \right).
   \end{split}
\end{equation*}

The Bhattacharyya loss function is therefore\footnote{To prevent infinite loss, we constrain $r_t$ to be at least $\frac{1}{\rm RTT}$ and at most $1 - \frac{1}{\rm RTT}$. While this may prevent the network from predicting absolute probabilities, it doesn't degrade performance in practice.},
\begin{equation}
\label{eqn:loss_B}
 \begin{split}
    \loss_B\paren{\{\hat{p}_j\}_{j=t^-+1}^t,\{b_j\}_{j=t^-+1}^t} = \\
    l_B \left( \frac{1}{\rm RTT}\sum_{j=t^-+1}^{t} \hat{p}_j, \frac{1}{\rm RTT} \sum_{j=t^-+1}^{t}b_j \right).
    \end{split}
\end{equation}

The training loss function is comprised of both losses, with hyperparameters  $\alpha, \beta>0$ balancing their contribution. The resulting training loss function is thus 
\begin{equation}
    \label{eqn:loss_MB}
    \loss_{MB} = \alpha\loss_{M} + \beta\loss_{B}.
\end{equation}

\subsection{ER-DeepNP}
\label{subsec:DeepNPBasic_acrlnc}
We propose embedding DeepNP with the \ac{acrlnc} scheme. At each time slot $t$, the feedback aggregated at the sender is used as input to the \ac{dnn} to generate the erasure rate estimation $\hat{\epsilon}_t$.
This estimation is then used by \ac{acrlnc} to determine the next transmission.
To better use these estimates, we suggest two modifications to the protocol presented in Subsection~\ref{subsec:Preliminaries}:
\begin{itemize}[wide, labelwidth=0.3cm, labelindent=1pt]
    \item The posterior \ac{fec} criterion is modified as follows, similar to \cite{cohen2022deepnp},     
    \begin{equation}
    \label{eqn:delta_new}
        \Delta_t \triangleq (md_t^{\text{nack}}+\epsilon_t \cdot c_{t}^{\text{new}}) - (ad_t^{\text{ack}}+r_t \cdot c_{t}^{\text{same}})-th_p,
    \end{equation}    
    where $md_t^{\text{nack}}$ and $ad_t^{\text{ack}}$ denote the DoFs with feedback acknowledgments that are missing or added, respectively. $c_{t}^{\text{new}}$ and $c_{t}^{\text{same}}$ denote the number of new and same coded packets transmitted in the unknown \ac{rtt} period, respectively. Hence, 
    $md_t^{\text{nack}}+\epsilon_t \cdot c_{t}^{\text{new}}$ expresses the amount of missing DoFs, while $ad_t^{\text{ack}}+r_t \cdot c_{t}^{\text{same}}$ the amount of added DoFs. Hence, re-transmission occurs when $\Delta_t > 0$, indicating excessive missing DoFs.
    
    \item Similar to \cite{michel2022flec,cohen2022broadcast}, the a-priori \ac{fec} mechanism is  time-dependent rather than data-packet dependent. Meaning, the sender transmits $\lceil \epsilon_t \cdot c_{t}^{\text{new}} \rfloor$ \ac{fec} every \ac{rtt}  slots.
\end{itemize}
The overall formulation of ER-DeepNP, which integrates DeepNP into \ac{acrlnc} as an erasure noise predictor, is summarized as Algorithm~\ref{alg:DeepNP}. The gray parts correspond to its extension to CL-DeepNP, detailed in Subsection~\ref{subsec:DeepNPExtended}.

\begin{algorithm}[t]\small
\caption{DeepNP}
\label{alg:DeepNP}
\begin{algorithmic}[1]
\State \textbf{Init:}
\State Trained DNN
\State AC-RLNC parameters - \(th\), \(w\), \(w_{\min}=1\), and \(w_{\max}=1\).
\State \colorbox{blue!15}{\parbox{\dimexpr\linewidth-3\fboxsep}{%
For continuous feedback: Initial \(S^{\rm th}\).
}}
\State \colorbox{gray!15}{\parbox{\dimexpr\linewidth-3\fboxsep}{%
\textbf{Extension to CL-DeepNP (Optional):}
Future rate determination slots: \(\tau\)
}}
\State \hrulefill

\State \textbf{Input:} Feedback \(F_{t^-}\).

\State \colorbox{blue!15}{\parbox{\dimexpr\linewidth-3\fboxsep}{%
For continuous feedback: Compute \(b_{t^-}\) using \eqref{eqn:b_t}.
}}

\State Stack \(F^{\text{in}}_t = [F_{t^--m+1},\ldots, F_{t^-}]\).

\Statex \colorbox{gray!15}{\parbox{\dimexpr\linewidth-4\fboxsep}{%
\textbf{Extension to CL-DeepNP (Optional):}
\State \hspace{0em} \textbf{if} \(t \bmod \tau == 0\) \textbf{then}
\State \hspace{1.5em} Determine \ac{mcs} index and \({\rm S_t^{\rm th}}\) via \ref{subsec:th_dec_p} or \ref{subsec:th_dec_deep}.
\State \hspace{0em} \textbf{endif}
}}
\Statex

\State \hspace{0em} \textbf{if} \(b_{t^-}=1\) \text{(Indicates decoding at the receiver)} \textbf{then} 
\State \hspace{1.5em} \(w_{\min} \gets\) next un-decoded data-packet.
\State \hspace{0em} \textbf{endif}

\vspace{0.2cm}
\Statex \hspace{0em} \underline{\textbf{EOW}}
\State \hspace{0em} \textbf{if} \(w_{\max} - w_{\min} > w\) \textbf{then}
\State \hspace{1.5em} Schedule a re-transmission.

\State \hspace{0em} \textbf{else}

\vspace{0.2cm}
\State \hspace{1.5em} \underline{\textbf{Noise prediction:}}
\State \hspace{1.5em} Feed \(F^{\text{in}}_t\) to DNN to get \(\{\hat{p}_j\}_{i=t^-+1}^t\).
\State \hspace{1.5em} Calculate \(\hat{\epsilon}_t\) using \eqref{eqn:SumSuccess}.

\vspace{0.2cm}
\Statex \hspace{1.5em} \underline{\textbf{A-Priori FEC}}
\State \hspace{1.5em} \textbf{if} \(t \bmod RTT == 0\) \textbf{then}
\State \hspace{3em} Schedule \(\lceil \hat{\epsilon}_t \cdot c_{t}^{\text{new}} \rfloor\) re-transmissions.

\State \hspace{1.5em} \textbf{else}

\vspace{0.2cm}
\State \hspace{3em} \underline{\textbf{Posterior FEC}}
\State \hspace{3em} Calculate \(\Delta_t\) using \eqref{eqn:delta_new}.

\State \hspace{3em} \textbf{if} \(\Delta_t > 0\) \textbf{then}
\State \hspace{4.5em} Schedule a re-transmission.
\State \hspace{3em} \textbf{else}

\vspace{0.2cm}
\State \hspace{4.5em} \underline{\textbf{New Coded Packet:}}
\State \hspace{4.5em} \(w_{\max} \gets w_{\max} + 1\).

\State \hspace{3em} \textbf{endif}
\State \hspace{1.5em} \textbf{endif}
\State \hspace{0em} \textbf{endif}

\State Generate new random coefficients \(\mu_i\) and encode:
\State \(c_{t}=\sum_{i=w_{\min}}^{w_{\max}}\mu_{i}P_{i}\) as given in \eqref{eqn:dof}.

\State \textbf{Output:} Next coded packet \(c_{t}\).
\end{algorithmic}
\end{algorithm}

\vspace{-0.4cm}
\subsection{CL-DeepNP}
\label{subsec:DeepNPExtended}
To balance the \ac{phy} layer code rate and the transport layer erasure rate, the sender can adjust the \ac{snr} threshold every few time slots. We next show how DeepNP can extended to CL-DeepNP, which facilitates intelligent threshold adjustments.

Consider \ac{phy} modulation with $U$ different coding options. Every $n$ time slots, the sender transmits packets with a certain \ac{phy} code rate and a corresponding threshold. Thus, for any $i \in [t, t+n]$, we can write $R_i^{\rm Phy} = R^{\rm Phy}(u)$ and  $S_i^{\rm th} = S^{\rm th}(u)$, where $u \in [1 \ldots U]$ is the chosen coding index. The erasure of packets sent in that time window, determined by \eqref{eqn:b_t}, may differ as a function of $u$ (via $S^{\rm th}(u)$). The estimated erasure probabilities may differ as well and can be written as,  $\{ \hat{p}_i(u)\}_{i=t}^{t+n}$. Thus, every $n$ time slots, the sender looks for $u^*$ that maximizes:
\begin{equation}
\label{eqn:max_th_decision}
    u^* = \argmax_{u \in U}\left(R^{\rm Phy}(u) \cdot \frac{1}{n} \sum_{i=t}^{t+n} \hat{p_i}(u)\right).
\end{equation}

\begin{figure}
\centering
  \includegraphics[width=0.85\linewidth,keepaspectratio]
  {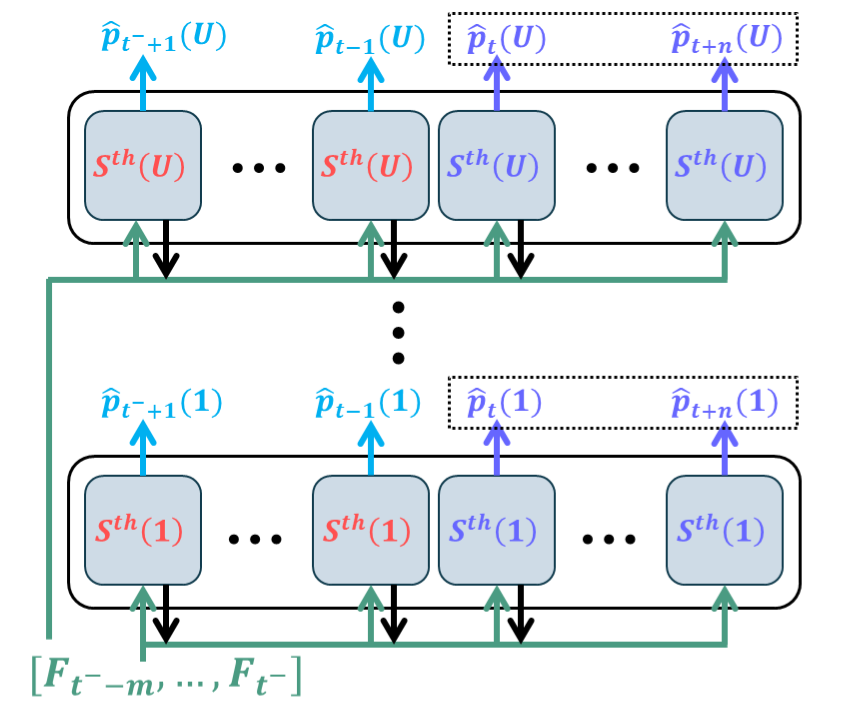}
  \caption{$U$ parallel DeepNP models extended to $\rm RTT$$+n$ building block. Each model has its own constant threshold.}
  \label{fig:dnn_th_par}
  \vspace{-0.4cm}
\end{figure}

\begin{figure}
\centering
  \includegraphics[width=\linewidth,keepaspectratio]
  {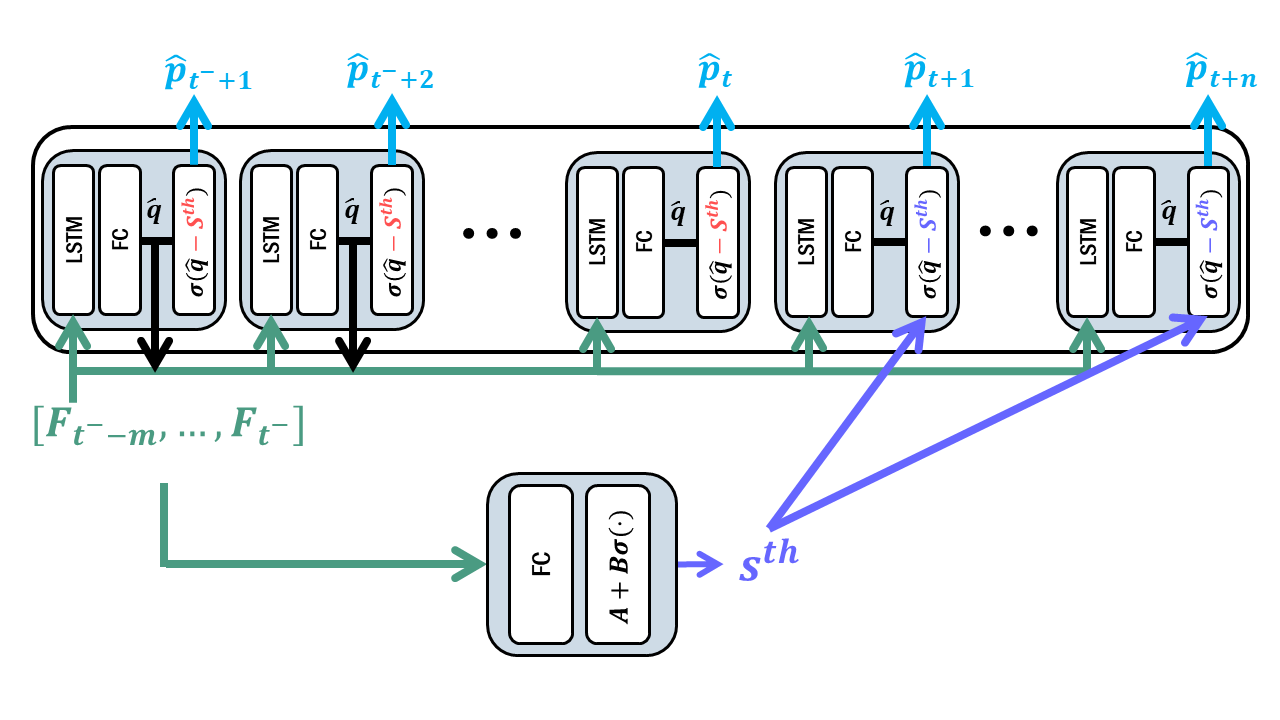}
  \caption{DeepNP  extended to $\rm RTT$$+n$ building blocks. The first $\rm RTT$ blocks' threshold is taken from the input, While the last $n$ blocks' threshold is generated from a separate neural block.}
  \label{fig:dnn_th_deep}
  \vspace{-0.4cm}
\end{figure}

Two thresholding approaches are suggested to tackle ~\eqref{eqn:max_th_decision}, realizing CL-DeepNP that jointly accounts for both the prediction of the erasure rate as well as the corresponding \ac{phy} modulation: 
\begin{enumerate}[wide, labelwidth=0.3cm, labelindent=1pt, label={T\arabic*}]
    \item \label{subsec:th_dec_p} In this approach, we train $U$ parallel DeepNP models, extended to $\rm {RTT} $$+ n$ neural building blocks. Each model has its own \ac{snr} threshold inside the sigmoid function- $\rm S^{\rm th}(u)$. During inference, every $\rm n$ time slots, the sender computes \eqref{eqn:max_th_decision} for every $u \in U$ and chooses the one that yields the maximum rate. This process is illustrated in Fig.~\ref{fig:dnn_th_par}.
    \item \label{subsec:th_dec_deep} Another learning-based solution enhances DeepNP to directly predict the threshold $\rm S^{\rm th}$ (without computing \eqref{eqn:max_th_decision}). This approach incorporates the threshold variability during transmission into its learning process. The modifications to DeepNP are presented in Fig.~\ref{fig:dnn_th_deep}.
\end{enumerate}

CL-DeepNP based on \ref{subsec:th_dec_p} can be viewed as using multiple ER-DeepNP algorithms as a form of an ensemble model~\cite{raviv2024crc}. However, the direct threshold setting in \ref{subsec:th_dec_deep} requires modifications to DeepNP as formulated in Subsections~\ref{subsec:DeepNPBasic_Arch}-\ref{subsec:DeepNPBasic_Train}. These modifications are,
\begin{enumerate}[wide, labelwidth=0.3cm, labelindent=1pt]
    \item {\bf Data} - The threshold used in the last \ac{rtt} transmissions is added to the input, i.e., 
    \begin{equation*}
            \mathcal{D} = \left\{
            \left( \{F_l\}_{l=t^--m+1}^{t^-}, 
            \{S_i^{\rm th} \}_{i=t^-+1}^{t} \right),
            \{b_j\}_{j=t^-+1}^t\right\}.
        \end{equation*}
    \item {\bf Architecture} - The number of neural blocks is extended to be ${\rm RTT}+n$. 
    The first ${\rm RTT}$ blocks represent previously transmitted packets. Hence, for any block $j$ such that $1 \leq j \leq {\rm RTT}$, the threshold in the sigmoid function is taken from the input, i.e., $\hat{p}_j = \sigma(\hat{q}_j -S^{\rm th}_{t^-+j})$.
    The last $n$ blocks correspond to unsent packets, allowing their threshold to be adjusted to future channel conditions. 
    A distinct trainable module generates a new threshold for these packets, comprising a \ac{fc} layer and a sigmoid function. 
    The sigmoid output, which is in the range $[0,1]$, is scaled such that the predicted $\hat{S}^{\rm th}$ satisfies 
    $\hat{S}^{\rm th} \in [ \min_{u \in U}{S^{\rm th}(u)}, \max_{u \in U}{S^{\rm th}(u)}]$.

    \item {\bf Loss} -
    Since we aim to maximize the code rate in learning, we need to represent its connection to the \ac{snr} threshold through a differentiable function. However, as detailed in Subsection~\ref{sec:sys_channel}, the \ac{phy} rate $R_t^{\rm Phy}$ is determined one-to-one with $\rm S_t^{\rm th}$ via the \ac{mcs}. This can be viewed as a piecewise constant function, whose gradient is zero almost anywhere. Following~\cite{shlezinger2022deep}, we use during training a {\em differentiable surrogate} of the \ac{phy} rate, approximating the piecewise constant function using shifted hyperbolic tangents. The resulting surrogate rate is computed as 
    \begin{align}    
        &\hat{R}^{\rm Phy}_{\rm smooth}( \hat{S}^{\rm th}) =  \, R(1) + \sum_{u=2}^U \left( \left( R^{\rm Phy}(u) - R^{\rm Phy}(u-1) \right)\right. \nonumber \\ 
        &\qquad\quad \cdot \frac{1}{2}\left.  \left(1 + \tanh\paren{b \cdot(\hat{S}^{\rm th} - S^{\rm th}(u)  ) }\right) \right).
        \label{eqn:r_smooth}
    \end{align}
    
    To learn the best threshold for maximizing \eqref{eqn:max_th_decision}, we subtract a rate-reward term from the  loss in \eqref{eqn:loss_MB} with hyperparameter $\gamma >0$, resulting in
    \begin{multline}
        \label{eqn:loss_th}
        \loss_{\rm th}\paren{\{\hat{p}_j\}_{j=t^-+1}^t, \hat{S}^{\rm th}, \{b_j\}_{j=t^-+1}^t} = \\
        \loss_{MB}\paren{\{\hat{p}_j\}_{j=t^-+1}^t,\{b_j\}_{j=t^-+1}^t} - \\ \gamma \cdot \hat{R}^{\rm Phy}_{\rm smooth} ( \hat{S}^{\rm th}) \cdot \frac{1}{n} \sum_{j=t}^{t+n} \hat{p_j}.
    \end{multline}

    \item {\bf Training} -
    We employ a two-step training process. The first step is a warm-up phase, where the primary objective is to train the model to accurately predict noise. During this phase, the loss function used is $\loss_{MB}$. In the second step, we fine-tune the warmed model by adjusting the thresholds, utilizing the loss function $\loss_{\rm th}$.
\end{enumerate}

\subsection{Discussion}\label{subsec:disc}

DeepNP is designed to facilitate noise prediction, as a means of boosting efficient network coding. 
It is particularly geared towards enhancing throughput-delay tradeoff (\ref{itm:tradeoff_T}) without requiring channel modeling (\ref{itm:Domain}), while exploiting non-binary feedback (\ref{itm:snr}) to boost performance and enable its cross-layer extension (\ref{itm:tradeoff_CL}).
This is achieved by employing the principled \ac{acrlnc} scheme while relaxing its reliance on channel modeling to predict the erasure rate.

Implementing noise prediction is challenging even when the underlying channel model is fully known, i.e., in the absence of \ref{itm:Domain}.
Commonly utilized approximations based on first and second-order statistical moments as in \eqref{e_est} are insufficient to reach optimal performance. 
DeepNP's architecture and training objectives are thus designed not to directly estimate the noise realization, but rather to meet the network code needs. Its architecture (Fig. \ref{fig:DeepNP_architecture}), which unrolls the noise prediction procedure over a short time window, allows the assignment of different weights for different time instances while exploiting its \acp{lstm} for tracking long correlations across time.
The training is based on a twofold objective goal - it jointly accounts for the fact that the \ac{dnn} yields individual noise predictions, while the network coding scheme requires an average erasure rate.
This approach enhances the network code's ability to mitigate the throughput-delay tradeoff (\ref{itm:tradeoff_T}).

The usage of \ac{lstm} building blocks facilitates exploiting the continuous nature of non-binary \ac{snr} feedback (\ref{itm:snr}).
It is particularly advantageous considering our focus on the threshold value rather than the exact \ac{snr} level. These factors significantly contribute to improved average erasure rate estimations, especially for larger \ac{rtt}s, as evidenced in Section \ref{sec:evaluation}. 
The \ac{snr} threshold chosen according to the \ac{mcs} serves as a tuning knob in determining the \ac{phy} code rate and the corresponding transport layer erasure rate (via \eqref{eqn:b_t}), thus meeting  \ref{itm:tradeoff_CL}. While the resulting CL-DeepNP requires some minor extensions in its architecture and training compared to the basic ER-DeepNP, it allows deviating from conventional layer-wise separate design to improve the overall performance, as demonstrated in Section~\ref{sec:evaluation}.

\rev{In the DeepNP architecture, all the operations in the LSTMs process are computationally light, consisting mainly of small matrix–vector multiplications. \textcolor{black}{In our implementation, running the trained DeepNP model using the publicly available code yields  latency values  comparable to typical URLLC slot durations, which can be further reduced using optimized inference libraries or dedicated hardware accelerators. This indicates on the ability of DeepNP to support real-time deployment within practical wireless systems. Still} this process may introduce some delay in updating the accurate estimated erasure rate for AC-RLNC decisions, which can be considered in longer RTT. However, it is important to note that: 1) the delay introduced by this architecture is mitigated by the size of the time slots for transmission in practical communication systems; thus, in practice, this delay is upper bounded by the duration of only a few time slots' transmissions. 2) The proposed solutions herein are much more robust to RTT variations, as evidenced in the simulation results in Section~\ref{sec:evaluation}. The impact of increasing the RTT time in a few time slots on the erasure rate estimation and the AC-RLNC protocol is illustrated by the evaluation study presented next.}
\rev{
In light of recent advances in transformer-based architectures for time-series modeling, we note that while these models excel at capturing very long-range dependencies, our choice of LSTM is motivated by its reduced computational and memory complexity, particularly important where the model is deployed as part of a real-time prediction mechanism. The underlying temporal dependencies are bounded by the RTT, resulting in a relatively short correlation horizon where the additional complexity of transformers would not yield substantial performance gains. We thus opted for LSTMs, a well-established architecture that effectively learns nonlinear temporal mappings with low computational overhead, aligning with the practical constraints of our framework.
Nevertheless, it would be interesting for future work to investigate whether modern efficient transformer variants could offer competitive performance while maintaining acceptable computational overhead for online deployment.
}

Furthermore, our DeepNP framework gives rise to a multitude of possible extensions. First, further investigations of the DeepNP with real-life data and systems stand as the next step to validate the solution's capability and resilience. Additionally, given the universal nature of DeepNP, it holds promise for application with other adaptive network coding schemes beyond \ac{acrlnc}, extending its potential utility. Another compelling avenue involves the incorporation of a noisy and quantized feedback channel, as well as the consideration of fluctuating \ac{rtt}. This could be implemented by leveraging cumulative feedback, as proposed in \cite{malak2019tiny}. Furthermore, the scope of  DeepNP could be broadened to encompass multi-path and multi-hop setups, thereby advancing its applicability in diverse network scenarios. These extensions are left for future study. 

\section{Performance Evaluation}
\label{sec:evaluation}
In this section, we evaluate our model using simulations\footnote{The source code, hyperparameters, and the data generator used in this experimental study are available at \url{https://github.com/Adinawx/SNR_DeepNP}.}. We begin with an overview of the simulation setup in Subsection~\ref{subsec:eval_exp_setup}, evaluating ER-DeepNP and CL-DeepNP in Subsections~\ref{subsec:eval_results}-\ref{subsec:CLDeepNP}, respectively.
\vspace{-0.2cm}
\subsection{Experimental Setup}\label{subsec:eval_exp_setup}
We use COST2100 \cite{cost2100}, a geometry-based stochastic channel model, to derive the channel's \ac{sinr}, which serves as a close approximation for the \ac{snr}. We use COST2100 to generate settings involving a \rev{Mobile Station (MS)}, communicating with a stationary \rev{Base Station (BS)}. To derive the \ac{sinr}, we generate four distinct taps. In each tap, the MS follows the same path with the same obstacles, while the BS is placed at a different location. One tap is designated as the primary transmission, while the other three represent interference and are multiplied by an attenuation factor.
The \ac{sinr} at time $t$ is then,
\begin{equation*}
    s_t = \frac{1}{\Omega}\sum_{\omega=1}^{\Omega}\left|\frac{H_0(t,\omega)}{H_1(t,\omega)+H_2(t,\omega)+H_3(t,\omega)}\right|^2 .
\end{equation*}
where $\Omega$ represents the number of transmitted frequencies in the chosen band, $H_0$ denotes the primary transmission, and $H_1$, $H_2$, and $H_3$ correspond to the interference taps.
For experiments on binary channels, we convert the \ac{sinr} series using a fixed threshold $S^{\rm th}$ and \eqref{eqn:b_t}.

\rev{
To analyze channels with different memory lengths, we explore three scenarios differing in the MS walking speed: Slow (velocity $\approx$ 0.2 m/s, representing a very slow-moving users such as a person standing or shifting position), Middle (velocity $\approx$ 0.5 m/s, representing pedestrian walking speeds), and Fast (velocity $\approx$ 1 m/s, representing quick walking). The Slow scenario reflects a long-memory channel characterized by infrequent and extended erasure bursts, while the Middle and Fast scenarios exhibit more frequent and shorter bursts due to increased temporal variation in the channel.
}
\rev{Each scenario comprises  a 3-second transmission duration with \ac{sinr} samples taken every 1 millisecond.
}
\textcolor{black}{The mobility regimes considered in this study (0.2--1 m/s) correspond to walking users and are intended to generate channel realizations with different temporal variability levels. In this context, the key factor affecting the prediction task is the rate of variation of the SINR over time rather than the absolute physical velocity. Consequently, these regimes should be interpreted as representing different temporal channel dynamics, which are conceptually equivalent to higher mobility scenarios with correspondingly denser sampling of channel observations. Furthermore, since the simulations rely on the COST2100 channel model, the resulting channel evolution is also influenced by environmental effects such as scattering clusters and visibility regions along the trajectory, which contribute to the observed SINR variability, not only by the terminal velocity.}

\begin{figure}
	\centering
         \includegraphics[width=0.5\textwidth, keepaspectratio]
         {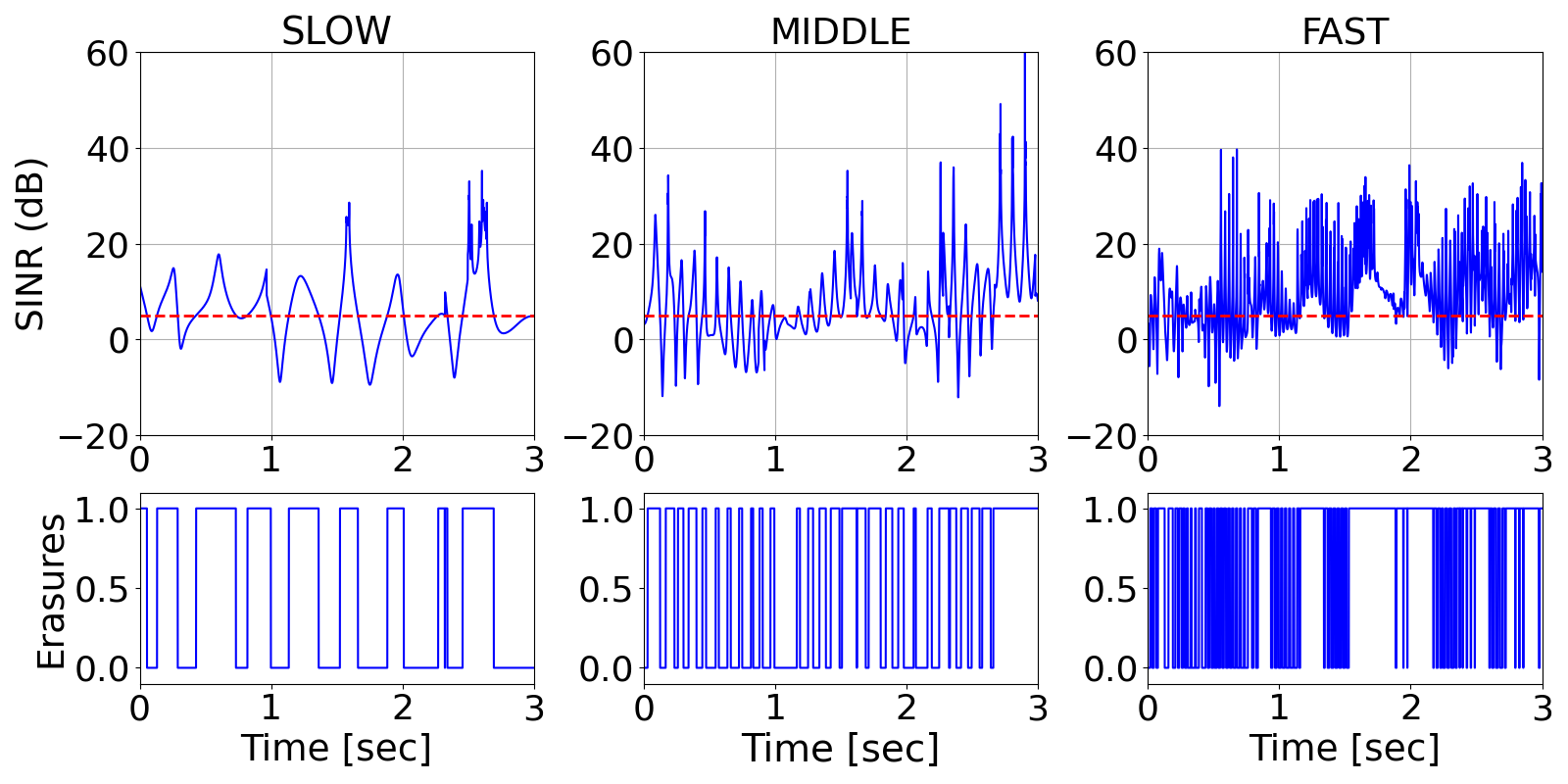}
         \vspace{-0.4cm}
        \caption{Simulation snapshot: visualizing SINR with a 5dB SINR threshold in the top row (red line), accompanied by the corresponding (binary) erasure series in the bottom row.}
	\label{fig:scenarios}
\end{figure}

Fig.~\ref{fig:scenarios} illustrates the outcomes of each scenario. The top three plots depict the \ac{sinr} in decibels, with the chosen \ac{sinr} threshold of 5dB. The bottom plot displays the corresponding erasure series. 
Table \ref{tab:scen_char} provides the corresponding details on the erasure burst amount, burst length, and erasure rate.
As expected, the {\em Slow} scenario, characterized by long memory, experiences a few long bursts. In contrast, the {\em Fast} scenario is unstable with numerous short bursts, while   {\em Middle} lies between these extremes.

\begin{table}
    \setlength{\tabcolsep}{0.2\tabcolsep}
    \centering
    \begin{tabular}{|m{2.3cm}|c|c|c|}
    \hline
         & SLOW & MIDDLE & FAST \\
    \hline
         Bursts number & $14.38 \pm 4.23$& $49.05 \pm 11.39$& $145.55 \pm 31.28$\\
    \hline
         Mean burst length & $136.2 \pm 64.7$& $39.36 \pm 13.36$& $13.51 \pm 3.8$\\
    \hline
         Max burst length & $710.82 \pm 377.8$& $361.44 \pm 240.91$& $221.17\pm 159.48$\\
    \hline
         Erasure rate & $0.34 \pm 0.16$ & $0.32 \pm 0.11$& $0.32 \pm 0.1$ \\
    \hline
    \end{tabular}
    \caption{Mean and standard deviation comparison for SINR threshold of $5$dB across all realizations.}
    \label{tab:scen_char}
    \vspace{-0.4cm}
\end{table}

We simulate the sender model in Fig.~\ref{fig:system_model}, which comprises three components: an erasure rate estimator, \ac{acrlnc} encoder, and a threshold decision block.
The \ac{sinr} is measured by the receiver and is sent to the sender via the feedback channel. Consequently, each slot in the transmission is equivalent to 1-millisecond of the generated \ac{sinr} series, and the total transmission is taken to be $T=2500$ time slots.

\rev{
Our DeepNP architecture employs a recurrent structure where each neural building block comprises an LSTM layer with 4 hidden units followed by a fully-connected (FC) layer with dimensions $4 \times 1$. The total number of blocks equals RTT for ER-DeepNP and $\text{RTT} + n$ for CL-DeepNP.
For training, we utilize the Adam optimizer \cite{kingma2014adam} with a learning rate of $0.01$, batch size of 300, and train for $100$ epochs with early stopping based on validation loss. The dataset comprises $1000$ Monte-Carlo simulations per scenario, with each simulation representing 3 seconds of transmission ($T = 2500$ time slots at 1ms granularity). We employ a $60-10-30$ split for training, validation, and testing sets respectively. The loss function hyperparameters are presented in the following subsections: for ER-DeepNP in Subsection \ref{subsec:eval_results} and for CL-DeepNP in Subsection \ref{subsec:CLDeepNP}. Models are trained separately for each mobility scenario (Slow, Middle, Fast) and for each feedback type (Binary/SINR).
Testing is performed across multiple RTT values: models trained for $\text{RTT} = 20$ are evaluated on $\text{RTT} = 10, 20$, while models trained for $\text{RTT} = 40$ are tested on $\text{RTT} = 30, 40$. The memory window size is set to $m = 4.5 \times \text{RTT}$. For CL-DeepNP-T2, we employ a two-stage training process as described in Section \ref{subsec:DeepNPExtended}.
}

\subsection{ER-DeepNP Results}\label{subsec:eval_results}

\rev{Here, we present the experimental results of the transport layer only, with the SINR threshold fixed at $5$dB}\footnote{\rev{In this numerical evaluation, we keep this threshold fixed for ER-DeepNP, as this approach is only affected by the erasure rate in the transport layer and does not count the MCS's coding rate of the PHY layer. For CL-DeepNP simulation results in Section~\ref{subsec:CLDeepNP}, this threshold varies to maximize \eqref{eqn:max_th_decision}.}}. 
 We consider the following erasure rate $\epsilon_t$ estimation options: 
    1) For a reference point (Ref), we take a "genie" model that is aware of the true channel rates. 
    2) A statistical model (Stat), taken as the mean over the available feedback, up to $m$ time slots back in time.
     This basic approach is an effective choice given the uncertainty of the channel model\footnote{\textcolor{black}{The statistical estimator represents the standard baseline used in AC-RLNC \cite{9076631,dias2023sliding,9245536,d2021post,cohen2022broadcast,vasudevan2026revisiting}, where the sender estimates the erasure rate from delayed feedback without assuming any channel model, consistent with the model-agnostic setting considered for DeepNP. More advanced predictors. (e.g., Kalman or autoregressive models) require parametric assumptions on channel dynamics, which are unavailable in our formulation and would require additional modeling considerations; therefore, the statistical estimator provides a fair baseline for comparison. The investigation of such predictors is left for future work.}}.
    3) ER-DeepNP with \ac{sinr} feedback (ER-DeepNP-SINR), and
    4) ER-DeepNP with Binary feedback (ER-DeepNP-Bin).


The loss function hyperparameters are chosen as $\alpha=1$ and $\beta=10$.

\subsubsection{Erasure Rate Prediction}
\begin{figure}
    \centering
    \includegraphics[width=\linewidth, keepaspectratio]{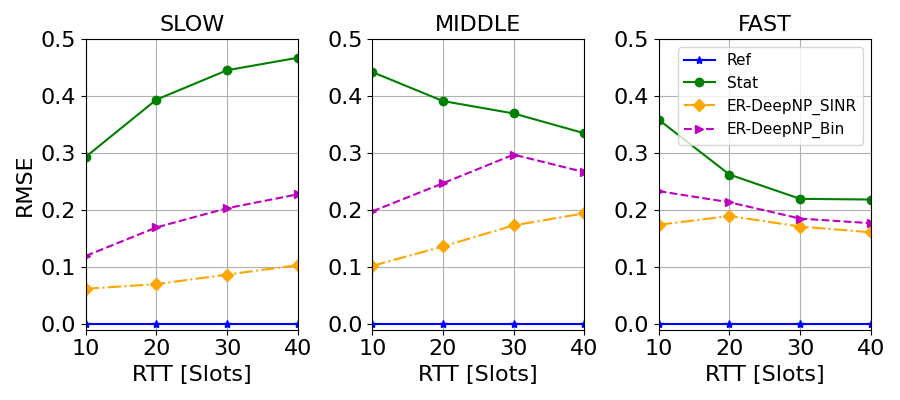}
    \caption{Channel rate prediction error.}
    \label{fig:rmse}
    \vspace{-0.4cm}
\end{figure}

We begin by evaluating DeepNP's accuracy separately from protocol performance, examining both the average and individual predicted erasure probabilities.
Unless mentioned otherwise, the results presented are for channels with erasure rates of $\epsilon \approx 0.3$.
\begin{figure}
    \centering
    \includegraphics[width=\linewidth,keepaspectratio]{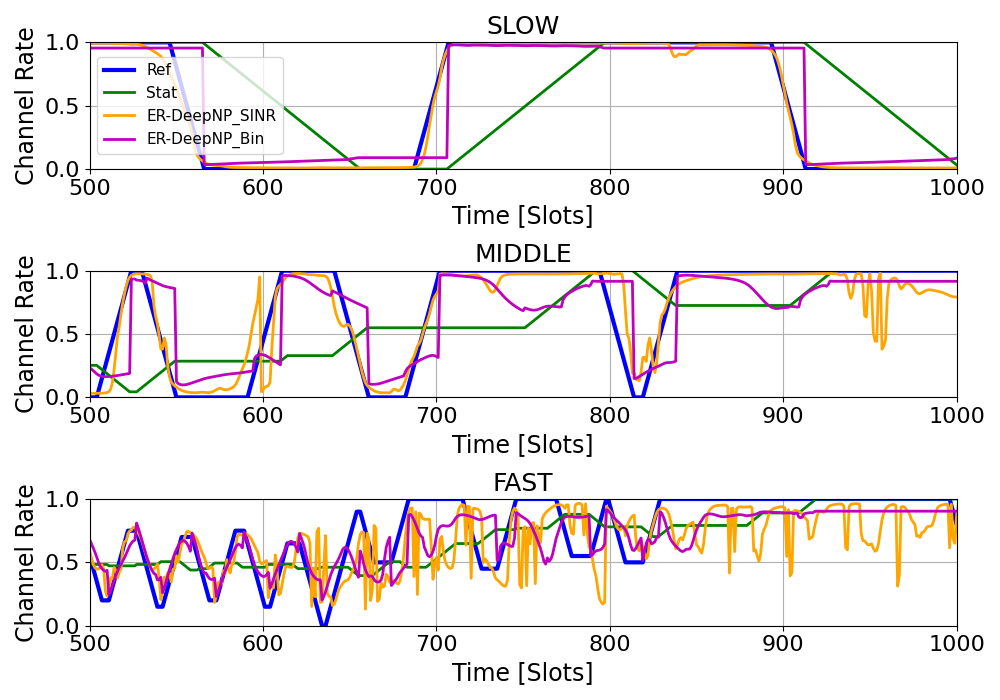}
    \caption{Transmission Snapshot: The predicted channel rate $r_t$ in each time slot for transmission with $\rm RTT = 20$.}
    \label{fig:ch_rate_real}
   \vspace{-0.3cm}
\end{figure}
Fig.~\ref{fig:ch_rate_real} illustrates the models' ability to predict the channel rate $r_t$. It presents a snapshot of one realization during time slots 500-1000 for $\rm RTT=20$, showing the predicted channel rates $r_t$ calculated in each slot for each scenario. The advantage of using the DeepNP models (SINR and Bin) over the statistical estimator is clear. The superiority of \ac{sinr} over the binary case is particularly evident in the transition areas when an erasure burst begins or ends.
%
Nonetheless, the capability of the proposed model to accurately track variations in the channel rate depends significantly on the rate of these variations. In more challenging scenarios - such as the {\em Fast} scenario - some oscillations occur when using soft SINR inputs instead of hard binary inputs. However, as shown next in Fig~\ref{fig:rmse}, its predictive capability directly translates into improved average channel rate prediction across a wide range of variation profiles and different \ac{rtt} lengths. These improvements, in turn, enhance the performance of the adaptive network coding, as demonstrated in the sequel. 
Hence, while one can potentially mitigate these variations by increasing the abstractness or modifications of training hyperparameters, Figs. ~\ref{fig:ch_rate_real} and ~\ref{fig:rmse} highlight the effectiveness of our approach in tracking channel rate fluctuations across diverse scenarios, ultimately strengthening adaptive network coding.

Fig.~\ref{fig:rmse} presents the RMSE between the approximated rate $\hat{r}_t$ and the actual rate $r_t$ (i.e., the RMSE between Ref and any other model in Fig. \ref{fig:ch_rate_real}). It reports the performance of various models tested over different \ac{rtt} for the three proposed channel conditions. This evaluation highlights the models' ability to estimate the channel rate over \ac{rtt} time windows, which is later important for protocol performance.
The statistical estimator errors in the {\em Middle} and {\em Fast} scenarios decrease for larger \ac{rtt}s, which can be attributed to their more frequent changes, causing a smaller variance for longer the input-output pairs (which are a function of \ac{rtt}). The {\em Slow} scenario, in contrast, exhibits an error increase, due to its more stable nature, which exhibits lower variance for shorter input-output windows.
While the DeepNP models might experience a slight degradation for larger \ac{rtt}s, they outperform the statistical estimator across all cases. This highlights their ability to capture the channel model, especially as channel memory lengthens, which is evident from the lower error in slower scenarios. Moreover, the utility of \ac{sinr} becomes apparent here, as the \ac{sinr}-based models surpass the binary models, particularly in the slow and middle scenarios.
\begin{figure}
    \centering
    \includegraphics[width=\linewidth, keepaspectratio]{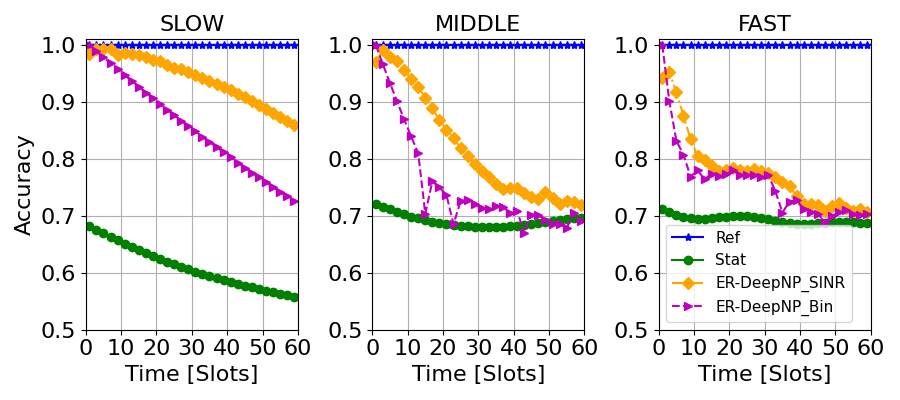}
    \caption{Prediction accuracy in each output slot.}
    \label{fig:StepAccuracy}
   \vspace{-0.4cm}
\end{figure}

While it is not crucial to the protocol performance, we are interested in the models' ability to accurately predict instantaneous erasure probabilities. Fig.~\ref{fig:StepAccuracy} shows the prediction accuracy for every second-time slot for $\rm RTT=40$ models. 
The predicted probability at each slot is rounded to 1 or 0, indicating reception or erasure, and compared with the actual erasure realization.
We observe in Fig.~\ref{fig:StepAccuracy} that the statistical estimator maintains approximately 0.7 accuracy in the {\em Middle} and {\em Fast} scenarios, which can be attributed to their more frequent changes, resulting in a lower variance across any time frame. On the contrary, the {\em Slow} case begins with 0.7 accuracy and degrades over time.
The DeepNP models consistently achieve greater accuracy compared to the statistical model, with the best prediction under the  {\em Slow} scenario, while approaching the statistical estimator over time as the channel memory diminishes in the {\em Fast} setting.
Again, The \ac{sinr} models show superiority over the binary models, especially in the slower scenarios.
The different losses exhibit only slight differences in specific time slots, which has minimal impact on the predicted rate. 

\subsubsection{Protocol Tests}
To evaluate the throughput-delay tradeoff, the protocol parameters were adjusted for each model to achieve comparable throughput levels, highlighting the tradeoff in the delay graphs.
\begin{figure}
    \begin{subfigure}{0.45\textwidth}
        \centering
        \includegraphics[width=\linewidth, keepaspectratio]
        {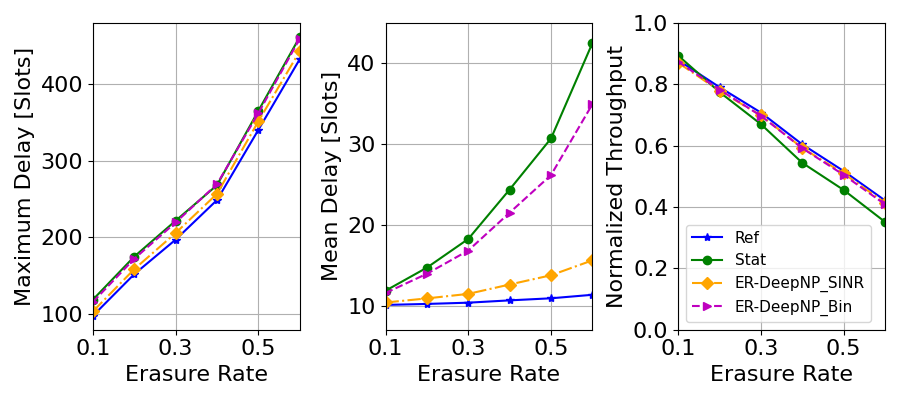}
        \caption{Slow scenario}
        \label{fig:protSlow_eps}
    \end{subfigure}
    \begin{subfigure}{0.45\textwidth}
        \centering
        \includegraphics[width=\linewidth, keepaspectratio]{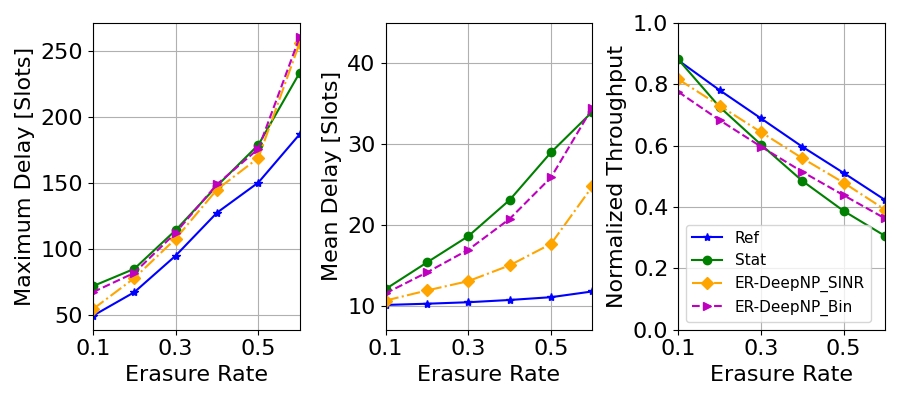}
        \caption{Middle scenario}
        \label{fig:protMid_eps}
    \end{subfigure}
    \begin{subfigure}{0.45\textwidth}
        \centering
        \includegraphics[width=\linewidth, keepaspectratio]{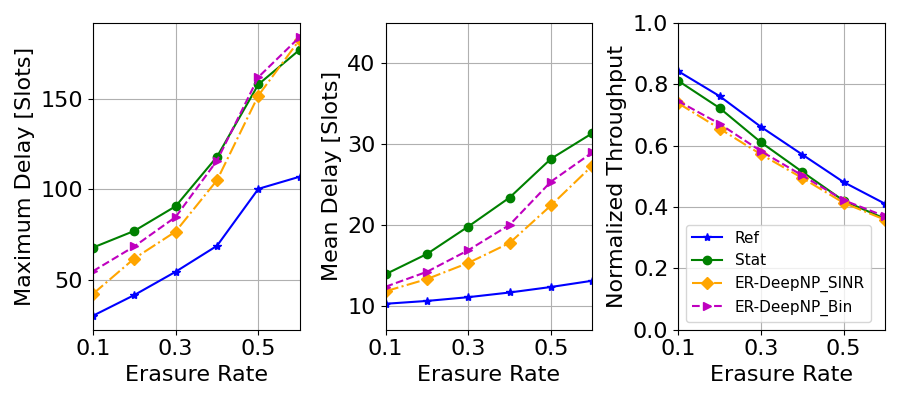}
        \caption{Fast scenario}
        \label{fig:protFast_eps}
    \end{subfigure}
    \caption{Protocol performance versus erasure rate, $\rm RTT = 20$.}
    \label{fig:prot_eps}
    \vspace{-0.4cm}
\end{figure}
Fig. \ref{fig:prot_eps} showcases the models' delay and throughput for $\rm RTT=20$ for increasing erasure rates.
The reference model achieves the optimal trade-off with throughput reaching capacity and a mean delay of $\frac{\rm RTT}{2}$ across all three scenarios. As longer channel memory and higher erasure rates result in longer erasure bursts, the maximum delay is inevitably higher for the slower scenarios. Additionally, it exhibits an increase across all scenarios as erasure rates increase.
The statistical estimator exhibits the longest delays in all scenarios and, as the channel's memory decreases, the models' performance is closer to the statistical approximations. 
Additionally, the \ac{sinr} models achieve a shorter delay for slow scenarios and are especially beneficial for higher erasure rates.


\begin{figure}
    \begin{subfigure}{0.45\textwidth}
        \centering
        \includegraphics[width=\linewidth, keepaspectratio]{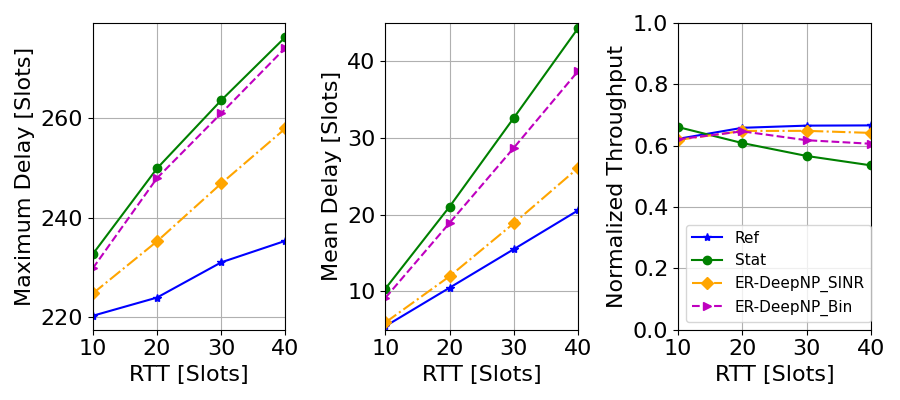}
        \caption{Slow scenario}
        \label{fig:protSlow_rtt}
    \end{subfigure}
    \begin{subfigure}{0.45\textwidth}
        \centering
        \includegraphics[width=\linewidth,keepaspectratio]{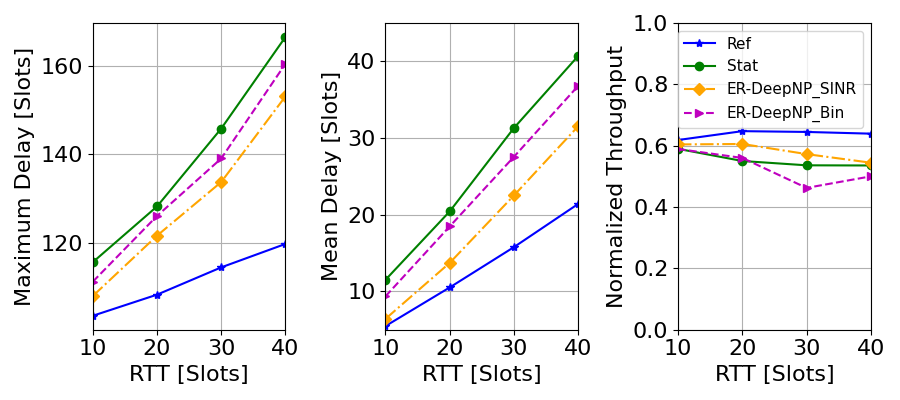}
        \caption{Middle scenario}
        \label{fig:protMid_rtt}
    \end{subfigure}
    \begin{subfigure}{0.45\textwidth}
        \centering
        \includegraphics[width=\linewidth, keepaspectratio]{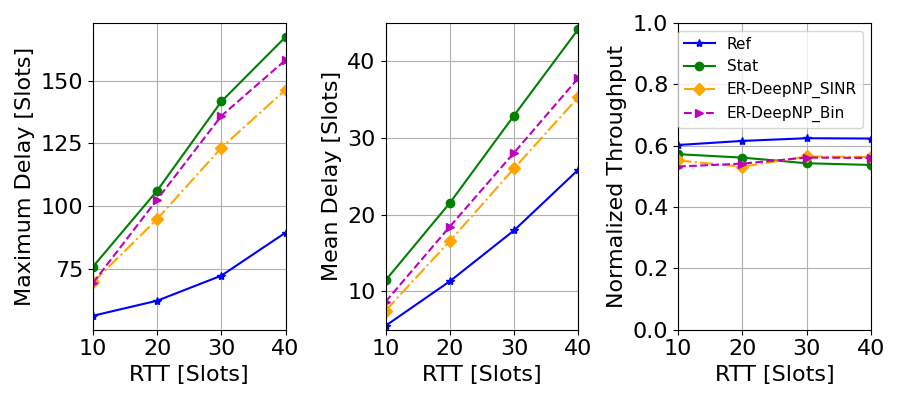}
        \caption{Fast scenario}
        \label{fig:protFast_rtt}
    \end{subfigure}
    \caption{Protocol performance versus RTT, erasure rate 0.3.}
    \label{fig:prot_rtt}
   \vspace{-0.4cm}
\end{figure}
Aligning with the prediction accuracy demonstrated in Fig.\ref{fig:rmse},  the delay and throughput are presented as a function of \ac{rtt} in Fig. \ref{fig:prot_rtt} for approximately $\epsilon=0.3$.
The reference model sets the optimal tradeoff, behaving similarly to Fig.~\ref{fig:prot_eps}.
The DeepNP models mitigate the tradeoff and achieve shorter delays compared to the statistical model, especially in slower cases.
For the presented erasure rate of $\epsilon=0.3$ the difference between using \ac{sinr} and binary feedback is evident in the {\em Slow} scenario. For greater erasure rates, the ability to leverage \ac{sinr} feedback becomes more notable, as observed in Fig.~\ref{fig:prot_eps}.

\subsubsection{Robustness}

\rev{To evaluate the generalization capability of DeepNP, we conduct two robustness studies. First, we examine cross-scenario performance, where models trained on one mobility pattern are tested on different one. Second, we assess the model's sensitivity to varying SINR levels.
}

\textit{3.1) Cross-Scenario Robustness:}
We evaluate the model's ability to generalize across different channel memory characteristics by testing models on mobility scenarios different from those used during training. Specifically, a model trained on data from the {\em Slow} scenario, is tested on the {\em Middle} dataset as well, denoted as "Same" and "Diff" (Different), respectively. This comparison assesses the model's capability to adapt from a bursty channel with long memory to a more arbitrary one. 
\begin{figure}
    \centering
    \includegraphics[width=0.9\linewidth]{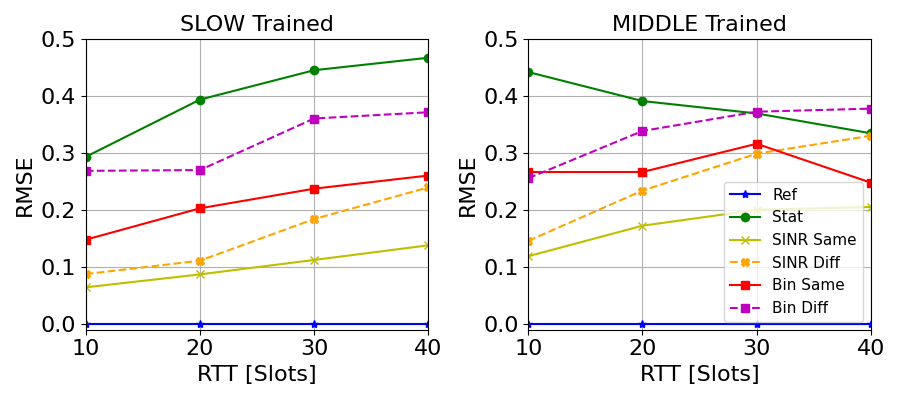}
    \caption{Robustness in channel rate prediction error. }
    \label{fig:rmse_Rob}
   \vspace{-0.4cm}
\end{figure}
In Fig.~\ref{fig:rmse_Rob} we see the error in the predicted channel rate $r_t$ as a function of \ac{rtt} in terms of RMSE, similar to Fig.~\ref{fig:rmse},
The errors in "Different" \ac{sinr} models increase with higher \ac{rtt}, yet they generally remain smaller than those of the statistical model. The error between the "Different" and "Same" models in the slow-trained model (left figure) is consistently smaller than in the middle-trained model (right figure). This illustrates that the slow-trained model, which specializes in long-term memory, can accurately predict shorter memory channels. In contrast, the middle-trained model focuses on detecting shorter-term patterns and struggles with longer memory channels over extended periods.
The "Different" binary models also outperform the statistical model in the {\em Slow} scenario, but approach the statistical model in the middle-trained model, demonstrating the efficiency of \ac{sinr} over binary data.
\begin{figure}
    \centering
    \includegraphics[width=0.9\linewidth]{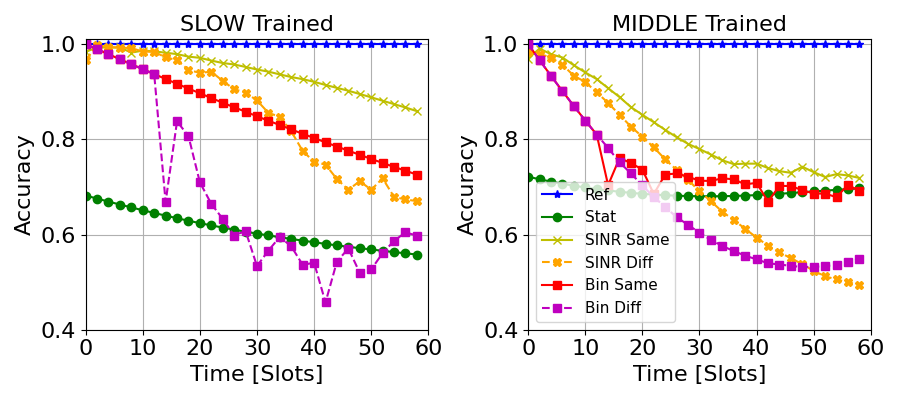}
    \caption{Robustness in prediction accuracy versus time slot.}
    \label{fig:StepAccuracy_Rob}
    \vspace{-0.4cm}
\end{figure}

In Fig.~\ref{fig:StepAccuracy_Rob}, we compare the model's ability to predict individual instantaneous erasures, similar to Fig.~\ref{fig:StepAccuracy}. 
Analyzing the \ac{sinr} models, the slow-trained model shows a degradation in the "Different" test compared to the "Same" model, but maintains overall high accuracy, surpassing the statistical estimator. The middle-trained "Different" model experiences a decline over time, exhibiting lower accuracy than the statistical estimator. This, once more, highlights the robustness of the slow-trained models compared to the middle-trained ones.
The binary models behave similarly to the \ac{sinr} models but show significantly greater degradation when comparing the "Different" and "Same" models in each scenario, especially in the slow-trained scenario. However, when compared to the statistical model, the slow-trained "Different" model performs better than the middle-trained one. This shows the robustness of \ac{sinr} over binary models and the effectiveness of long-memory learning against channel changes.

\begin{figure}
    \begin{subfigure}{0.45\textwidth}
        \centering
        \includegraphics[width=\linewidth, keepaspectratio]{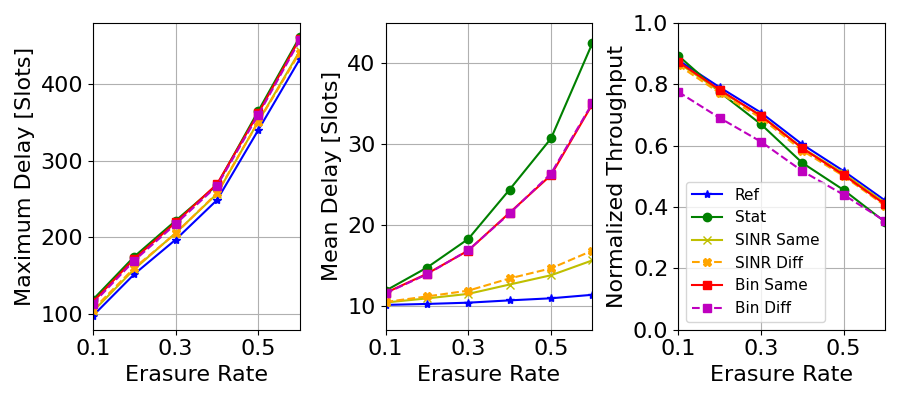}
        \caption{Slow trained model}
        \label{fig:protSlow_eps_rob}
    \end{subfigure}
    \begin{subfigure}{0.45\textwidth}
        \centering
        \includegraphics[width=\linewidth, keepaspectratio]{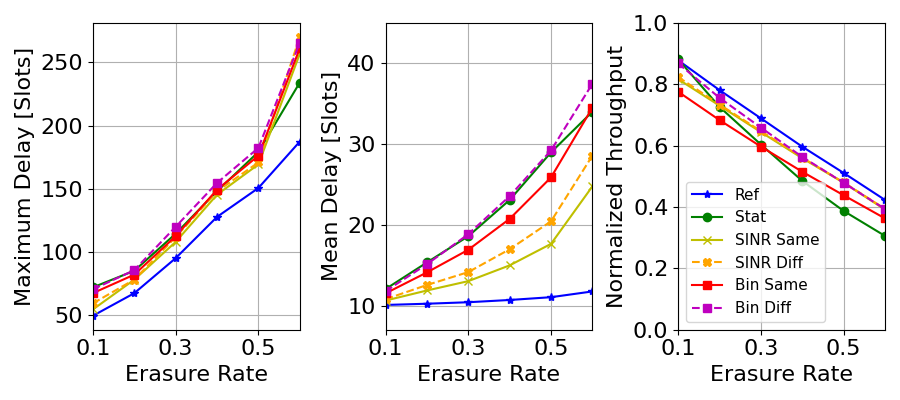}
        \caption{Middle trained model}
        \label{fig:protMid_eps_rob}
    \end{subfigure}
    \caption{Robustness performance versus  erasure rate.}
    \label{fig:prot_eps_rob}
    \vspace{-0.4cm}
\end{figure}

\begin{figure}
    \begin{subfigure}{0.45\textwidth}
        \centering
        \includegraphics[width=\linewidth, keepaspectratio]{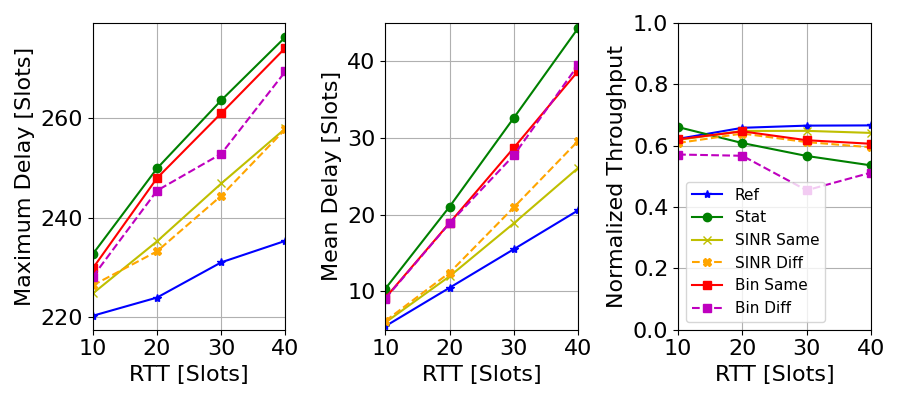}
        \caption{Slow trained model}
        \label{fig:protSlow_rtt_rob}
    \end{subfigure}
    \begin{subfigure}{0.45\textwidth}
        \centering
        \includegraphics[width=\linewidth, keepaspectratio]{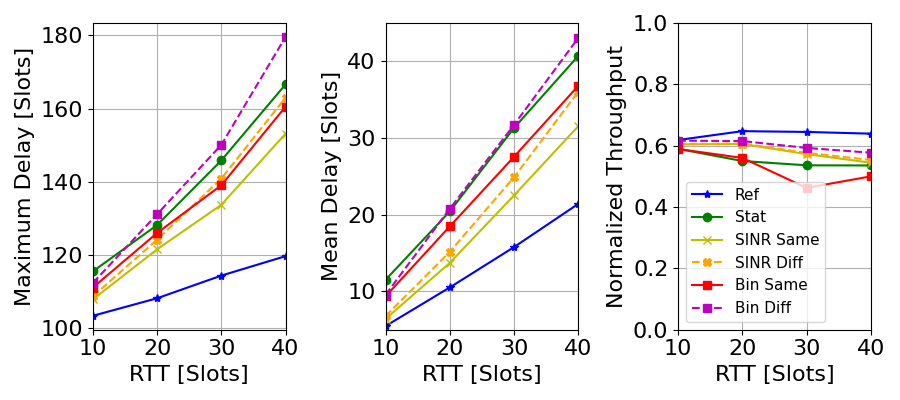}
        \caption{Middle trained model}
        \label{fig:protMid_rtt_rob}
    \end{subfigure}
    \caption{Robustness performance versus RTT.}
    \label{fig:prot_rtt_rob}
\end{figure}
Robustness is evident when running the protocol in Figs. \ref{fig:prot_eps_rob} and \ref{fig:prot_rtt_rob}, presenting the throughput and delay versus erasure rate and \ac{rtt} as done in Figs.~\ref{fig:prot_eps} and \ref{fig:prot_rtt} respectively. 
The "Different" \ac{sinr} consistently shows slightly higher delays compared to the "Same" models, as observed in the accuracy results, yet it generally achieves lower delays than the statistical model. Similarly, in the middle-trained case, the binary "Different" models occasionally converge to the statistical model or are slightly worse. However, these differences are minimal, emphasizing the overall robustness of DeepNP.

{\color{black}
\textit{3.2) Robustness to Varying SINR Levels:}
To assess robustness across diverse channel conditions, we evaluate our methods beyond the baseline mean SINR of $\gamma \approx 7$\,dB used in previous subsections. Since wireless environments experience signal quality variations due to distance, mobility, and environmental factors, we test DeepNP across mean SINR values from $\gamma \approx 3$\,dB to $\gamma \approx 6$\,dB by increasing the white Gaussian noise in the simulations.
Decreasing the SINR level in this manner reduces the burstiness of the instantaneous SINR signal and alters the channel's temporal characteristics. This phenomenon is evident in the results presented below.

Figures~\ref{fig:rmseRobNoise_rtt}--\ref{fig:prot_eps_RobNoise} compare performance across different mean SINR levels. Solid lines represent testing on channels matching the training condition ($\gamma \approx 7\,\text{dB}$), while dashed and dotted lines show performance on channels with mean SINR of 4\,dB and 6\,dB, respectively. Each curve is shown for its corresponding model: reference model (blue), statistical model (green), and ER-DeepNP model (yellow).

\begin{figure}
    \centering
    \includegraphics[width=\linewidth, keepaspectratio]{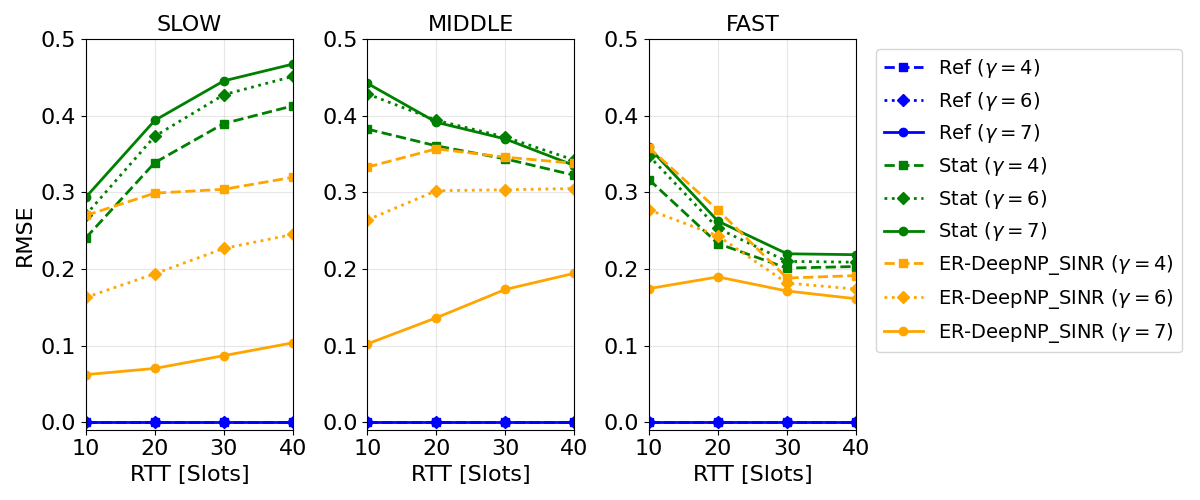}
    \caption{Channel rate prediction error versus RTT for different mean SINR level. }
    \label{fig:rmseRobNoise_rtt}
    \vspace{-0.3cm}
\end{figure}

Fig.~\ref{fig:rmseRobNoise_rtt} presents the RMSE between the predicted rate $\hat{r}_t$ and the actual rate $r_t$ as a function of RTT. 
In the \textit{Slow} scenario, the trained model maintains low error at 6\,dB. At 4\,dB, performance degrades, but DeepNP still outperforms the statistical model, particularly at higher RTTs. Interestingly, the statistical models perform relatively better at lower SINR levels where channel characteristics become more stochastic - a trend observed across all scenarios.
In the \textit{Fast} scenario, performance converges toward the statistical estimator as SINR decreases due to increasingly arbitrary signal behavior, consistent with Subsection~\ref{subsec:eval_results}. This convergence demonstrates a form of robustness - the model degrades to statistical performance rather than producing larger errors.
The \textit{Middle} scenario performance falls between the \textit{Slow} and \textit{Fast} cases.

\begin{figure}
    \begin{subfigure}{0.45\textwidth}
        \centering
        \includegraphics[width=\linewidth, keepaspectratio]{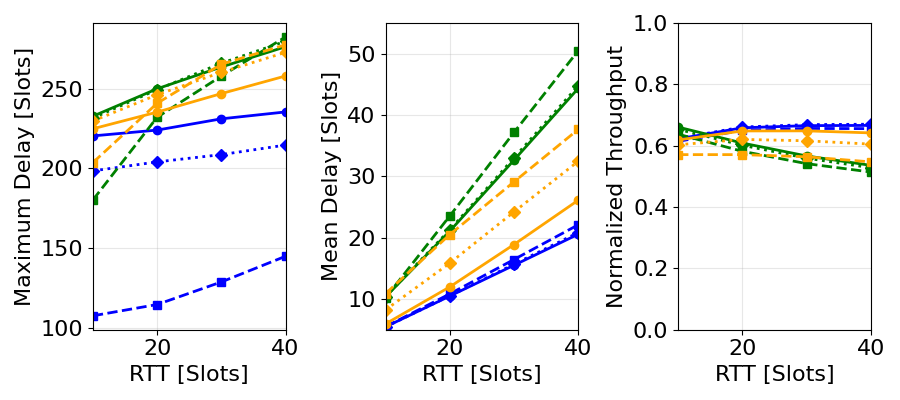}
        \caption{Slow scenario}
        \label{fig:protSlow_rtt_RobNoise}
    \end{subfigure}
    \begin{subfigure}{0.45\textwidth}
        \centering
        \includegraphics[width=\linewidth,keepaspectratio]{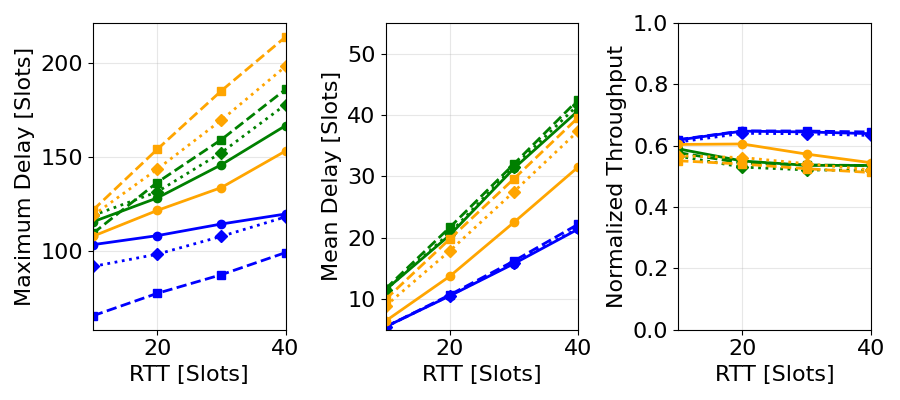}
        \caption{Middle scenario}
        \label{fig:protMid_rtt_RobNoise}
    \end{subfigure}
    \begin{subfigure}{0.45\textwidth}
        \centering
        \includegraphics[width=\linewidth, keepaspectratio]{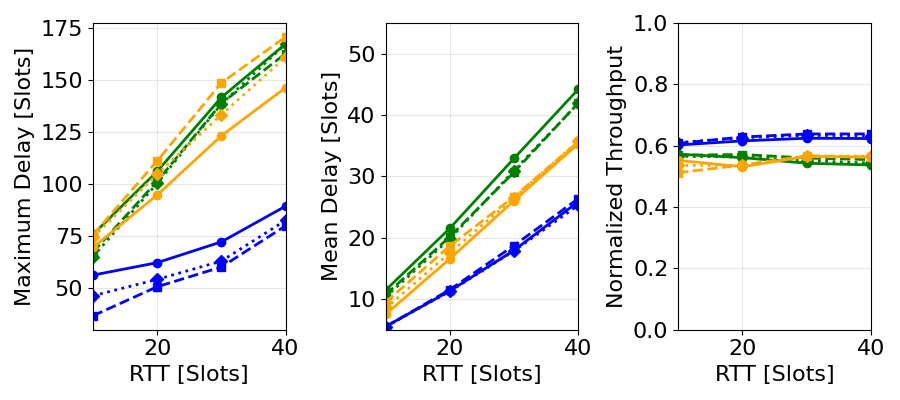}
        \caption{Fast scenario}
        \label{fig:protFast_rtt_RobNoise}
    \end{subfigure}
    \caption{Protocol performance versus RTT for different mean SINR levels. Erasure rates is 0.3 for all models. The legend is given in Fig.~\ref{fig:rmseRobNoise_rtt}.}
    \label{fig:prot_rtt_RobNoise}
   \vspace{-0.4cm}
\end{figure}

\begin{figure}
    \begin{subfigure}{0.45\textwidth}
        \centering
        \includegraphics[width=\linewidth, keepaspectratio]{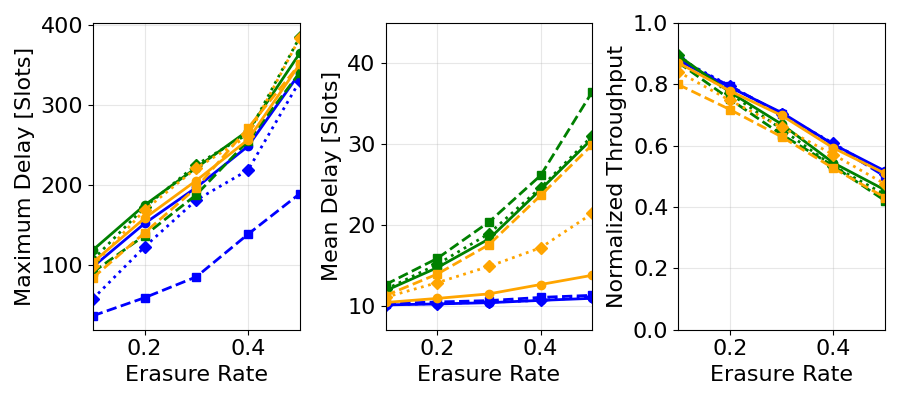}
        \caption{Slow scenario}
        \label{fig:protSlow_eps_RobNoise}
    \end{subfigure}
    \begin{subfigure}{0.45\textwidth}
        \centering
        \includegraphics[width=\linewidth,keepaspectratio]{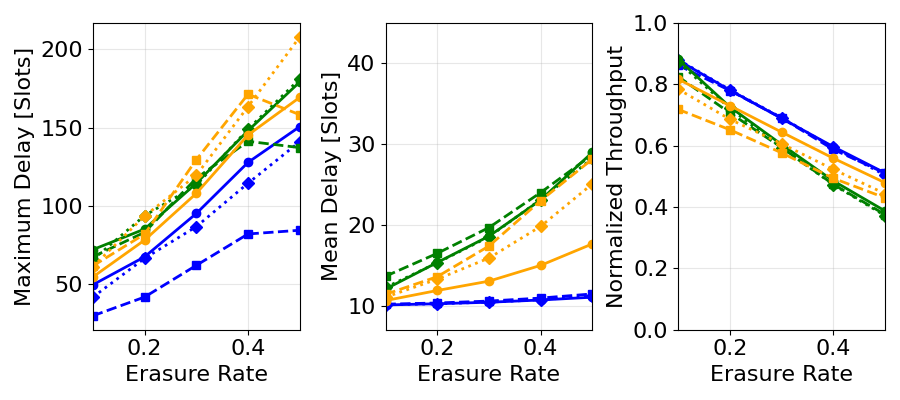}
        \caption{Middle scenario}
        \label{fig:protMid_eps_RobNoise}
    \end{subfigure}
    \begin{subfigure}{0.45\textwidth}
        \centering
        \includegraphics[width=\linewidth, keepaspectratio]{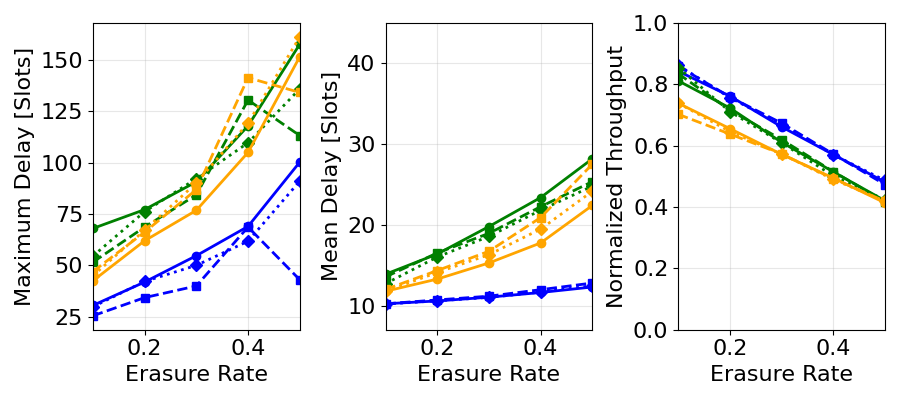}
        \caption{Fast scenario}
        \label{fig:protFast_eps_RobNoise}
    \end{subfigure}
    \caption{Protocol performance versus erasure rate for different mean SINR levels, RTT=20. The legend is given in Fig.~\ref{fig:rmseRobNoise_rtt}.}
    \label{fig:prot_eps_RobNoise}
   \vspace{-0.4cm}
\end{figure}
Figures~\ref{fig:prot_rtt_RobNoise}-~\ref{fig:prot_eps_RobNoise} present the protocol performance metrics - normalized throughput, mean delay, and maximum delay - as functions of RTT and erasure rate, respectively, for the three SINR conditions. 
As in previous sections, we focus on delay performance since the throughput is similar across different conditions. Consistent with RMSE results (Fig.~\ref{fig:rmseRobNoise_rtt}), the \textit{Slow} scenario shows increased mean delay as SINR decreases, while the \textit{Fast} scenario exhibits a similar delay across SINR levels, outperforming the statistical model. The \textit{Middle} scenario falls between these extremes.
The maximum delay reveals a distinctive phenomenon: very low SINR levels (4\,dB) result in lower maximum delay across all scenarios as the added noise causes more frequent but shorter erasure bursts, as opposed to the fewer and longer bursts observed at moderate SINR levels (6–7\,dB).
All these trends remain consistent when examining performance as a function of erasure rate, as seen in Figure~\ref{fig:prot_eps_RobNoise}.

\begin{figure}
    \centering
    \includegraphics[width=\linewidth, keepaspectratio]{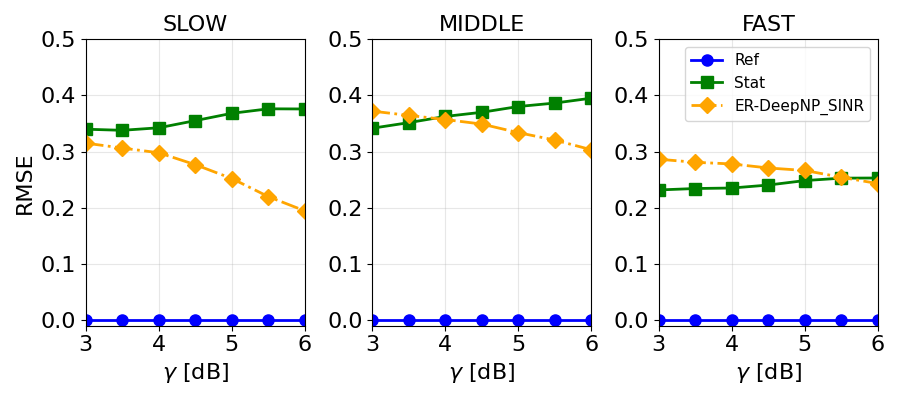}
    \caption{Channel rate prediction error versus mean SINR level at RTT=20. }
    \label{fig:rmseRobNoise_noise}
    \vspace{-0.4cm}
\end{figure}

\begin{figure}
    \begin{subfigure}{0.45\textwidth}
        \centering
        \includegraphics[width=\linewidth, keepaspectratio]{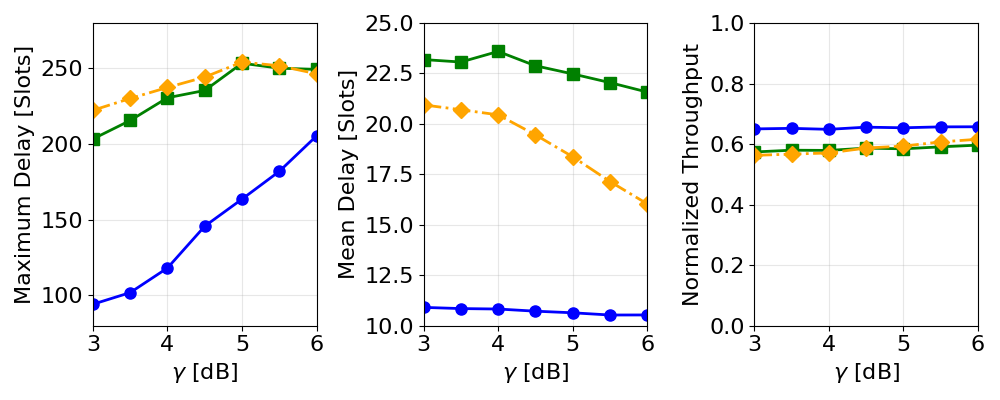}
        \caption{Slow scenario}
        \label{fig:protSlow_RobNoise}
    \end{subfigure}
    \begin{subfigure}{0.45\textwidth}
        \centering
        \includegraphics[width=\linewidth,keepaspectratio]{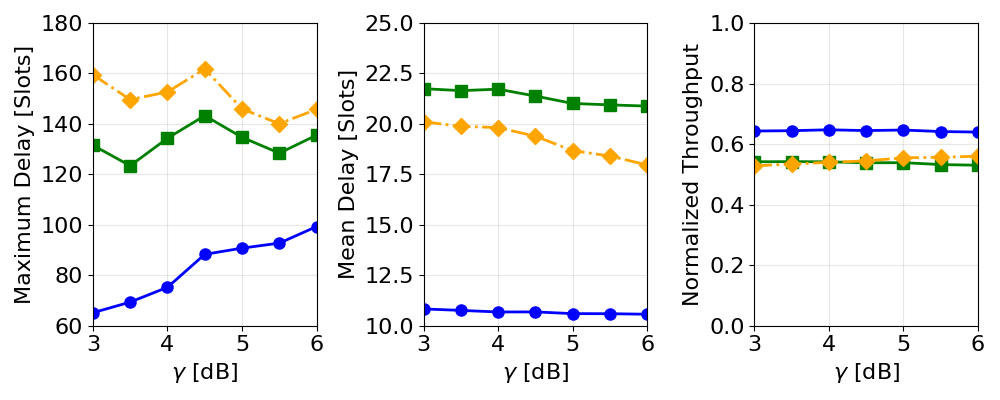}
        \caption{Middle scenario}
        \label{fig:protMid_RobNoise}
    \end{subfigure}
    \begin{subfigure}{0.45\textwidth}
        \centering
        \includegraphics[width=\linewidth, keepaspectratio]{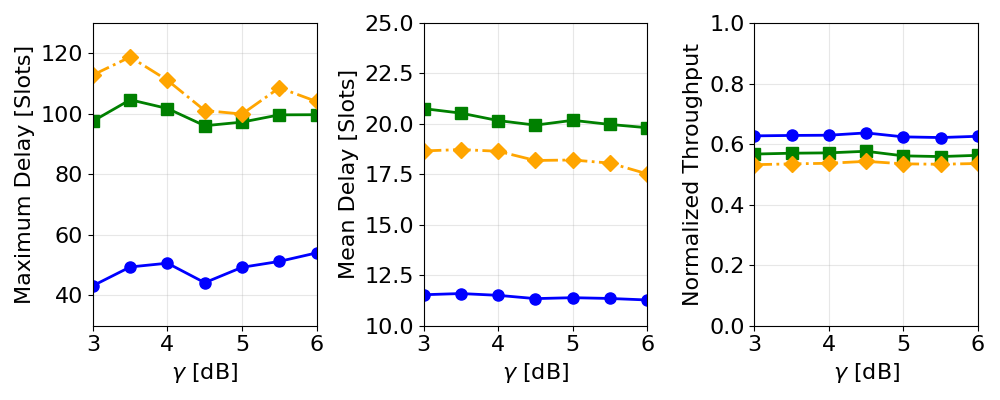}
        \caption{Fast scenario}
        \label{fig:protFast_RobNoise}
    \end{subfigure}
    \caption{Protocol performance versus mean SINR level at RTT=20. The legend is given in Fig.~\ref{fig:rmseRobNoise_noise}.}
    \label{fig:prot_RobNoise}
   \vspace{-0.4cm}
\end{figure}

Figures~\ref{fig:rmseRobNoise_noise}-\ref{fig:prot_RobNoise} present results as a function of SINR level (3–6\,dB) for RTT=20. Blue solid lines represent the reference model, green solid lines the statistical model, and dashed yellow lines the ER-DeepNP model.
The \textit{Slow} scenario shows improved performance with increasing mean SINR level, maintaining advantages over the statistical baseline across the entire range. The \textit{Middle} scenario matches statistical performance at 4\,dB and improves at higher SINR levels. The \textit{Fast} scenario demonstrates robustness from 5\,dB onward, consistent with results from Subsection~\ref{subsec:eval_results} and Fig.~\ref{fig:rmseRobNoise_rtt}.
Protocol performance (Fig.~\ref{fig:prot_RobNoise}) reflects these prediction trends for mean delay, particularly in the \textit{Slow} scenario. These results demonstrate a robustness range of 2-3\,dB depending on the mobility scenario.

}
\rev{
Overall, these robustness experiments demonstrate that while DeepNP performance degrades outside its training regime, it maintains advantages over statistical estimation, particularly for longer-memory channels (\textit{Slow} scenario). The demonstrated 3\,dB SINR level robustness range and adaptability across mobility scenarios make DeepNP suitable for real-world systems. These evaluations effectively test adaptation to abrupt changes in channel memory and erasure patterns - extreme forms of environmental variation. In practical systems experiencing time-varying conditions, periodic retraining could maintain optimal performance as channel statistics evolve.
}

\vspace{-0.4cm}
\subsection{CL-DeepNP Results}
\label{subsec:CLDeepNP}
We proceed to evaluate CL-DeepNP, evaluating our cross-layer design. 
We use the rate-threshold lookup table in Table \ref{tab:mcs},  where the values are chosen to match our data and are similar to the \ac{mcs} given in \cite[Page 34]{ccs_specs}. We compare the following models.  
The first two models are used to demonstrate the importance of using an adaptive threshold. Accordingly, we present results using a predefined constant threshold for
1) a reference model (Const Ref)  and 
2) ER-DeepNP. Both are taken for a $5$dB threshold, which results in an erasure rate similar to the other models, allowing for a fair comparison.
These are compared to adaptive thresholds using:
3) a reference model (Ref) and
4) a statistical model (Stat) that determines their threshold similar to the parallel models' option presented in \ref{subsec:DeepNPExtended}, by computing \eqref{eqn:max_th_decision}.
5) CL-DeepNP model using \ref{subsec:th_dec_p} approach (CL-DeepNP-T1), trained with $\alpha=1$ and $\beta=10$.
6) CL-DeepNP using \ref{subsec:th_dec_deep} (CL-DeepNP-T2), with $\alpha=1$, $\beta=50$, and $\gamma=100$ for the loss function \eqref{eqn:loss_th} and a smooth factor of $b^{-1}=0.7$ in \eqref{eqn:r_smooth}.
The CL-DeepNP models are trained and tested similarly to the ER-DeepNP models, with an extension of $n = \frac{\rm RTT}{2}$ blocks.

 \begin{figure}
    \centering
    \includegraphics[width=0.55\linewidth]{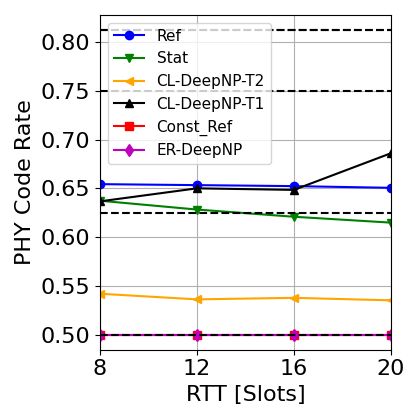}
    \caption{Physical layer code rate. The dashed lines represent the optional code rates corresponding to the MCS Table~\ref{tab:mcs}.}
    \label{fig:phy_rate}
\end{figure}
Fig. \ref{fig:phy_rate} presents the mean code rate chosen across the entire transmission of $T=2500$ time slots.
The constant models' code rate is, of course, stable at the $0.5$. The adaptive statistical and adaptive reference models lie close to $0.625$, with the CL-DeepNP-T1 as well, which only experiences a slight increase for \ac{rtt}=20. CL-DeepNP-T2 model is in-between, approximately at $0.55$.

\begin{figure}
\vspace{-0.3cm}
        \centering
        \includegraphics[width=\linewidth]{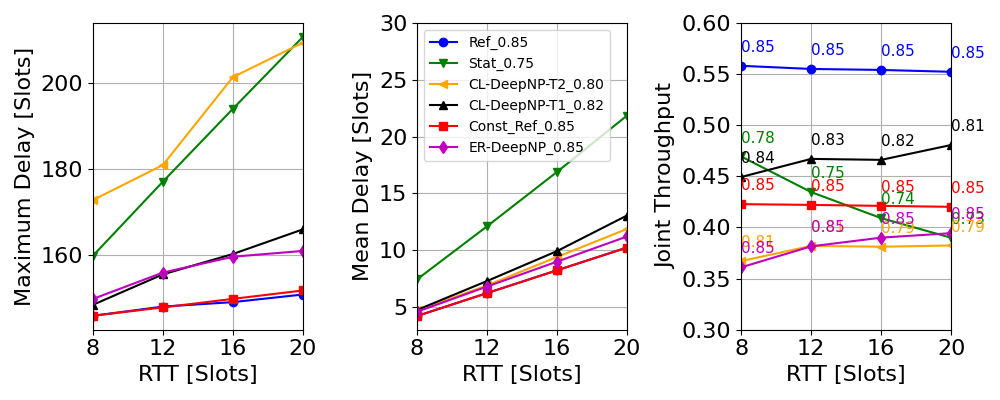}
        \label{fig:protSlow}
        \vspace{-0.4cm}
    \caption{Protocol performance versus RTT, Slow scenario.}
    \label{fig:prot_rate}
    \vspace{-0.4cm}
\end{figure}

To understand this behavior, we report the joint-throughput-delay tradeoff in Fig.~\ref{fig:prot_rate}
The figure presents the joint throughput in \eqref{eqn:eta_j} along with maximum and mean delay. Since alternating the threshold changes the erasure rate, each point in the throughput graph is presented with a note of the resultant average channel rate $r$, as well as in the mean rate of all \ac{rtt}s in the legend.
All models reach a nice mean delay close to $\frac{\rm RTT}{2}$, except the statistical model which experiences a higher delay, similar to the transport mean delay in Fig.~\ref{fig:prot_rtt}. 
The maximum delay varies respectively with the changing channel rates. 
The advantage of threshold adjustments is evident when comparing the adaptive reference (Ref) and constant reference (Ref Const) models. For a similar channel rate, the adaptive reference achieves a higher joint throughput while maintaining similar delay levels.
Analyzing the CL-DeepNP models, \ref{subsec:th_dec_p} approach (of parallel trained models) outperforms the adaptive statistical model. It also achieves higher joint throughput compared to the other models, and its maximum delay is similar to that of the constant deep model. Overall, this demonstrates the benefits of this approach.
The \ref{subsec:th_dec_deep} approach (deep threshold learning) reaches an intermediate level of performance. Compared to the constant deep model (ER-DeepNP), it experiences more erasure bursts, resulting in a higher maximum delay. However, at a lower channel rate, it exhibits the same joint throughput. Although this model's results are not optimal, they showcase the potential of this method. Further investigation is suggested to modify the learning process to achieve smaller bursts and lower maximum delay.

\section{Conclusions}\label{sec:conclusions}
In this study, we introduced DeepNP, a novel approach that utilizes data-driven learning with adaptive network coding to predict noise rate without the need for explicit channel modeling. 
DeepNP, designed to predict instantaneous channel noise and estimate erasure rates,  is demonstrated to enhance \ac{acrlnc} performance over burst channels, showcasing notable improvements in both throughput and delay simultaneously.
By leveraging \ac{snr} values, DeepNP maintains high performance even with large round-trip times and high erasure rates. Additionally, the integration of \ac{phy} layer indications allows for the creation of a cross-layer DeepNP variant. This variant balances the \ac{phy} layer code rate with the transport layer erasure rate by tuning the \ac{snr} limit of the \ac{mcs} during transmission. As a result, it achieves higher throughput across the network.

\bibliographystyle{IEEEtran}
\bibliography{IEEEabrv, bibtex/references,bibtex/Ref1,bibtex/Ref2}

\end{document}